\documentclass[prl,twocolumn,amssymb,floatfix]{revtex4-1}
\usepackage{amsmath} 
\usepackage{amsfonts} 
\usepackage{amssymb}
\usepackage{bbm}
\usepackage{epsfig}
\usepackage{graphics}
\usepackage{graphicx}

\def\ptps{{Prog.\ Theor.\ Phys.\ Suppl. \ }}

\newcommand{\lsim}{\,\lower2truept\hbox{${<\atop\hbox{\raise4truept\hbox{$\sim$}}}$}\,}
\newcommand{\gsim}{\,\lower2truept\hbox{${>\atop\hbox{\raise4truept\hbox{$\sim$}}}$}\,}
\newcommand{\pp}{~~~.}
\newcommand{\vv}{~~~,}

\newcommand{\be}{\begin{equation}}
\newcommand{\ee}{\end{equation}}
\newcommand{\bea}{\begin{eqnarray}}
\newcommand{\eea}{\end{eqnarray}}
\newcommand{\beann}{\begin{eqnarray*}}
\newcommand{\eeann}{\end{eqnarray*}}
\newcommand{\nn}{\nonumber}

\newcommand{\scalprod}[2]{\mathbf{#1 \hspace{-.2em}\cdot\hspace{-.2em} #2}}
\newcommand{\exvalue}[1]{\langle #1 \rangle}

\begin{document}

\title[Neutrino lumps and the Cosmic Microwave Background]
{Neutrino lumps and the Cosmic Microwave Background}

\author{Valeria Pettorino$^{1}$, Nico Wintergerst$^2$, Luca Amendola$^{3}$, Christof Wetterich$^3$}
\affiliation{
$^1$ SISSA, Via Bonomea 265, 34136 Trieste, Italy,
\\
$^2$ Arnold-Sommerfeld-Center, Ludwig-Maximilians-Universit\"at, Theresienstr. 37, D-80333 M\"unchen, Germany,
\\
$^3$ Institut f\"ur Theoretische Physik, Universit\"at Heidelberg,
Philosophenweg 16, D-69120 Heidelberg, Germany.
}

\begin{abstract}
  The interaction between the cosmon and neutrinos may solve the ``why now problem'' for dark energy cosmologies. Within growing neutrino quintessence it leads to the formation of nonlinear neutrino lumps. For a test of such models by the integrated Sachs-Wolfe effect for the cosmic microwave background (CMB) we estimate the size and time evolution of the gravitational potential induced by these lumps. A population of lumps with size of $100$ Mpc or more could lead to observable effects on the CMB anisotropies for low angular momenta. The linear approximation is found to be invalid for the relevant length scales. Quantitative estimates depend strongly on the details of the transition between the linear and nonlinear regimes. In particular, important backreaction effects arise from the nonlinearities of the cosmon interactions. At the present stage the uncertainties of the estimate make it difficult to constrain the parameter space of growing neutrino models. We explicitly discuss scenarios and models that are compatible with the CMB observations.
\end{abstract}

\date{\today}

\maketitle

\section{Introduction}

Quintessence models \cite{wetterich_1988,ratra_peebles_1988} with a growing neutrino mass can explain why dark energy dominates in the present cosmological epoch \cite{amendola_etal_2007,wetterich_2007}. A crucial ingredient is the coupling of the cosmon to neutrinos which results in an interaction between dark energy and matter \cite{wetterich_1995,amendola_2000}. Indeed, the cosmon-neutrino coupling $\beta$ can be large (as compared to gravitational strength) \cite{fardon_etal_2004,bjaelde_etal_2007,brookfield_etal_2005}, such that even the small fraction of cosmic energy density in neutrinos can have a substantial influence on the time evolution of the quintessence field or cosmon. This effect becomes important only once the neutrino mass plays a role. The transition from relativistic to nonrelativistic neutrinos at redshift $z_{\text{NR}} \sim 5 - 10$ therefore constitutes a ``cosmological trigger event'' which stops the further time evolution of the cosmon. For $z < z_\text{NR}$ the time evolution of the cosmic scale factor behaves very similar to an effective cosmological constant which becomes computable in terms of the present day neutrino mass \cite{amendola_etal_2007, wetterich_2007}.

For realistic cosmologies the cosmon-neutrino coupling is larger than one, corresponding to an enhancement $\sim \beta^2$ of the cosmon-mediated attraction between nonrelativistic neutrinos as compared to their gravitational attraction. In turn, this substantially accelerates the growth of neutrino structures for $z < z_\text{NR}$ - the characteristic timescale is shortened by a factor $\beta^{-2}$. Large scale stable neutrino lumps \cite{brouzakis_etal_2007, bernardini_etal_2009} can form after a redshift $z_\text{nl} \approx 1-2$ at which the neutrino fluctuations become nonlinear \cite{mota_etal_2008, wintergerst_etal_2009}. The gravitational potential induced by neutrino lumps can have important effects on the propagation of photons and therefore influence the CMB-anisotropies via the integrated Sachs-Wolfe (ISW) effect. An extrapolation of the linear fluctuation analysis in presence of the strong neutrino-cosmon coupling would indeed rule out most such models \cite{afshordi_etal_2005, franca_etal_2009}. 

However, the linear analysis becomes invalid for the relevant length or momentum scales. An extrapolation of the linear evolution violates rapidly even the most extreme bound on the neutrino induced gravitational potential which would result if all neutrinos of the visible universe are concentrated in one spot \cite{wintergerst_etal_2009}. The use of such an extrapolation for validity tests and parameter estimates of specific growing neutrino models \cite{franca_etal_2009} can yield misleading results. The challenge to be solved is an estimate of the neutrino-lump induced cosmological gravitational potential on large scales. 

The size of this potential and its time history determines the size of the ISW effect which may be observed by a modification of the CMB-spectrum at low angular momentum $l$. It also influences the correlation between matter and photon fluctuations for which the present observations yield values larger than expected in the cosmological $\Lambda$CDM model \cite{jain_etal_1998, inoue_etal_2006, rudnick_etal_2007, samal_etal_2007, giannantonio_etal_2008}. Finally, a rather sudden increase of the neutrino-induced gravitational potential could reconcile \cite{ayaita_etal_2009} the presently observed large bulk flows \cite{watkins_2008} with observational bounds on the matter fluctuations at similar scales \cite{reid_etal_2009}. If the attractive force between fluctuations is only mediated by gravity rather tight bounds relate the growth rate and the peculiar velocities to the density contrast. This bound gets weakened if the time scale is shortened, as characteristic for the nongravitational cosmon-mediated force.

The difficulty of a reliable estimate of the neutrino-induced gravitational potential comes from the simple observation that a gas of neutrino lumps behaves differently from a fluid of individual neutrinos. This concerns both gravitational and cosmon-mediated effects. On small enough length scales, where neutrino lumps form first, the nonlinear part of the evolution is dominated by the formation of individual lumps. We have computed the time history of individual lumps by solving the relevant nonlinear Navier-Stokes equations in the Newtonian limit \cite{wintergerst_etal_2009}. The local gravitational potential on a length scale characteristic for the lump increases faster than in the linear approximation during an ``infall period''. Subsequently, it settles to an almost constant value. With reasonable assumptions about the distribution of the lumps these results can be used for an estimate of the neutrino induced cosmological potential $\Phi_\nu(k)$. The neutrino fluctuations in this ``nonlinear range'' of rather large $k$ have no substantial effect on the CMB.

On the other hand, for length scales sufficiently large compared to the horizon, linear perturbation theory applies. Again, the neutrino fluctuations in the ``linear regime'' have only minor effects on the CMB. Visible effects can arise from neutrino induced fluctuations of the gravitational potential $\Phi_\nu(k)$ in the intermediate range of scales, typically only moderately inside the horizon, which interpolates between the linear and nonlinear ranges. A potential $\Phi_\nu(k)$ exceeding substantially a value around $10^{-5}$ would have a large impact on the CMB and rule out the corresponding model. 

The gravitational side of the breakdown of linear perturbation theory for neutrino fluctuations resembles in some aspects the situation for cold dark matter fluctuations, but also shows important differences. The latter is well understood. Cold dark matter fluctuations grow nonlinear first on scales much smaller than the horizon. For example, galaxies are formed long before the emergence of superclusters, happening in the present cosmological epoch. The complicated nonlinear formation process of a galaxy has only little impact on its gravitational potential at a distance $l$ if the infall of additional matter into a volume $l^3$ around the galaxy can be neglected. Once the infall of new matter stops, the gravitational potential of a single galaxy for a characteristic scale $l$ does not change in time anymore. The cosmological gravitational potential (more precisely the root of the squared fluctuations of the potential in Fourier space) at a comoving wave number $k = \pi a/l$ changes then dominantly due to the expansion of the universe and decreases due to dilution. 

In principle, the formation of galaxies could influence the growth of structure on length scales much larger than galactic scales. Beyond linear perturbation theory the fluctuations with different $k$ are coupled such that a formal expansion in lowest order of the fluctuations breaks down once the galactic-scale fluctuations grow nonlinear. In principle, a gas of galaxies needs not to be the same as a fluid of dark matter particles. This extends to larger nonlinear cosmic structures like clusters or local voids. For the case of gravity, however, the question if structure formation can influence the cosmological evolution has been answered to the negative \cite{wetterich_2003}. This applies similarly to the evolution of fluctuations for $k$ much smaller than the characteristic $k_\text{nl}$ where fluctuations have grown nonlinear. The central point of this argument states that the gravitational field sufficiently outside a mass concentration or energy density inhomogeneity is quite insensitive to the details of the mass distribution or local relative velocities. The gravitational field equation is linear in the energy momentum tensor such that nonlinear ``back-reaction'' effects can only arise due to self-interactions of the gravitational field. For realistic cosmological structures a typical estimate of the gravitational backreaction yields relative effects only of the order $10^{-5}$ \cite{wetterich_2003, brown_etal_2009, behrend_etal_2008}. In short, linear perturbation theory for fluctuations at a scale $k$ is not affected by nonlinearities at larger $k' \gg k$, provided that the gravitational potential $\Phi(k')$ induced by these ``local lumps'' remains small compared to one. (Challenges of this simple picture by speculations about large gravitational backreaction effects \cite{kolb_etal_2005} appear not to be compatible with a realistic distribution of structures in the universe). The decoupling of the small distance nonlinearities from the linear evolution of large distance inhomogeneities is the basis for the predictions of the CMB-anisotropies in the $\Lambda$CDM model.

For neutrino fluctuations the gravitational aspects of the difference between a gas of neutrino lumps and a fluid of neutrino particles differ from cold dark matter in two important aspects. The first concerns the difference in the relevant length scales. For cold dark matter the scales of substantial nonlinearities do not exceed much the scales of clusters and the nonlinear Rees-Sciama effect affects the CMB only for angular momenta $l > 1500$ \cite{rees_sciama_1968}. Due to the additional cosmon-mediated attractive force, the neutrino fluctuations become nonlinear for much smaller $k$ as compared to dark matter. Nonlinear lumps with size exceeding considerably $100$ Mpc may form. This implies that the breakdown of linear perturbation theory affects the CMB-anisotropies for much smaller angular momenta $l$. For the models that we discuss in this paper the linear approximation for the CMB breaks down for $l \le 200$; for $k > 10$ Mpc$^{-1}$ the neutrino fluctuations remain linear due to the effect of free streaming.

The second difference is related to the dependence of the neutrino mass on the value of the cosmon field. Its value inside the lumps may differ substantially from the average cosmological value which would be computed for an almost homogeneous neutrino fluid. Since the mass is the source of the gravitational potential at all scales, a concentration of a substantial fraction of neutrinos within lumps where their mass differs from the cosmological value could result in important modifications for the evolution of the gravitational potential on large length scales due to nonlinearities on small scales. This possibility of a large backreaction effect differs from pure gravity, where the particle masses are constant.

The main differences between a neutrino fluid and a gas of neutrino lumps concern the cosmon field rather than the gravitational aspects. For growing neutrino quintessence the dominant attractive force between nonrelativistic neutrinos is mediated by the cosmon. In a Newtonian approximation the role of the gravitational potential is taken by $\beta\,\delta\phi$, with $\delta\phi$ the local fluctuation of the cosmon or quintessence field. While the gravitational potential remains much smaller than one even for highly nonlinear neutrino lumps, the quantity $\beta\,\delta\phi$ can reach values of order one, signalling a breakdown of linear theory in the cosmon sector. Indeed, in a Newtonian approximation $\beta\,\delta\phi$ exceeds the gravitational potential by a factor $2\beta^2$ which reaches values $10^4$ - $10^6$ in our models. A local value $\beta\,\delta\phi = 1$ means that the local mass of the neutrinos is smaller by a factor $e^{-1}$ as compared to the cosmological neutrino mass. One expects substantial modifications of linear theory once $\beta\,\delta\phi$ reaches values of order unity.

In the presence of a cosmon field the impact of small scale structure on the behavior of fluctuations at larger scales (backreaction) is generically expected to be much larger than in the case of gravity \cite{wetterich_2003,schrempp_brown_2009}. The reason are the nonlinearities in the cosmon field equations. They concern both the role of the cosmon potential and cosmon-neutrino coupling, in particular if the effective interaction strength $\beta(\phi)$ depends on the value of the cosmon field. As mentioned above, these nonlinearities may cause a difference between the average neutrino mass inside the neutrino lumps as compared to the value which would be obtained for a roughly homogeneous fluid of neutrinos. Also the effective cosmon-neutrino coupling $\beta = -\partial\ln{m_\nu}/\partial\phi$ may be reduced for a gas of neutrino lumps as compared to a homogeneous neutrino fluid. 

In this paper we account for the effect of the nonlinearities by simple physically motivated bounds on the growth of fluctuations. This will allow us to give a first estimate of the effects of neutrino lumps on the CMB anisotropies. Our findings clearly indicate that linear perturbation theory is not appropriate. On the other hand, the uncertainties of the quantitative treatment of the nonlinear effects remain large, such that reliable estimates of the CMB spectrum are not yet available for growing neutrino quintessence. The latter would require a better quantitative estimate of the cosmological gravitational potential induced by neutrino lumps, perhaps by numerical N-body simulations or nonperturbative analytic methods.

\section{Growing neutrino quintessence} \label{gnm}

The time evolution of growing neutrino models follows a set of equations \cite{mota_etal_2008} which combines the evolution of the homogeneous and isotropic background with the growth of perturbations. We recall for convenience the essential ingredients characterizing these models.
At the background level, the expansion of the universe obeys the Friedmann equation:
\be \label{f1} {\cal H}^2 \equiv \left(\frac{a'}{a}\right)^2 = \frac{a^2}{3} \sum_\alpha \rho_\alpha \ee
where primes denote derivatives with respect to conformal time $\tau$. The sum is taken over all components $\alpha$ of the energy density in the universe, including dark matter, quintessence, neutrinos, baryons and radiation. The time evolution of the energy density $\rho_\alpha$ for each species involves the equation of state $w_\alpha \equiv p_\alpha/\rho_\alpha$.

A crucial ingredient in this model is the dependence of the average neutrino mass on the cosmon field $\phi$, as encoded in the dimensionless cosmon-neutrino coupling $\beta$, \be \label{beta_phi} \beta \equiv - \frac{d \ln{m_\nu}}{d \phi} \pp \ee For increasing $\phi$ and constant $\beta < 0$ the neutrino mass increases with time \be \label{eq:nu_mass} m_{\nu} = \bar{m}_{\nu} e^{-{{\beta}} \phi} \vv \ee where $\bar{m}_{\nu}$ is a constant. In general, $\beta$ can be a function of $\phi$, as proposed in \cite{wetterich_2007} within a particle physics model, modifying equation (\ref{eq:nu_mass}) but leading to similar effects. The cosmon field $\phi$ is normalized in units of the reduced Planck mass $M = (8 \pi G_N)^{-1/2}$. For a given
cosmological model with a given time dependence of $\phi$, one can determine
the time dependence of the neutrino mass $m_\nu(t)$. For three degenerate
neutrinos the present average value of the neutrino mass $m_\nu(t_0)$ can be related
to the energy fraction in neutrinos ($h \approx 0.72$) \be \label{omeganu} \Omega_{\nu} (t_0) = \frac{3 m_\nu
  (t_0)}{94\, eV h^2} \,\, . \ee 

The cosmon exchange mediates an additional attractive force between neutrinos of strength $\beta^2$. The case $\beta \sim 1$ corresponds to a strength comparable to gravity. In this paper we investigate two models, one with constant $\beta = -275$ and $m_\nu(t_0) = 0.47$ eV, and 
the other with varying $\beta(\phi)$ as proposed in \cite{wetterich_2007}. In this ``growing coupling model'' an approach towards zero of the large mass scale in a see-saw type mechanism for the neutrino mass is reflected by a strong increase
of the light neutrino masses that we take all equal for simplicity,
\be 
m_\nu = \frac{\bar{m}_\nu \phi_t}{\phi_t - \phi}   \,\,\,\,\,\,\,  , \,\,\,\,\, \beta = - \frac{1}{\phi_t - \phi} \,\,\,\,\,\, , \,\,\,\,\,\, \frac{\partial \beta}
{\partial \phi} = - \beta^2 \vv
\ee
with constants $\bar{m}_\nu$ and $\phi_t$. In a coupled model this behavior is only approximate. The pole at $\phi = \phi_t$ is not reached in practice. We choose a present neutrino mass $m_\nu(t_0) = 0.46$ eV and $\beta(t_0) = -150$.

The dynamics of the cosmon can be inferred from the Klein
Gordon equation, now including an extra source due to the neutrino coupling,

\be \label{kg} \phi'' + 2{\cal H} \phi' + a^2 \frac{dU}{d \phi} = a^2 \beta
(\rho_{\nu}-3 p_{\nu}) \,\, , \ee with $\rho_\nu$ and $p_\nu = w_\nu \rho_\nu$ 
the energy density and pressure of the neutrinos. We choose
an exponential potential \cite{wetterich_1988, ratra_peebles_1988, ferreira_joyce_1998, barreiro_etal_2000, wetterich_2007}:

\be \label{pot_def} V(\phi) = M^2 U(\phi) = M^4 e^{- \alpha \phi} \vv \ee
where the constant $\alpha$ is one of the free parameters of our model and determines the amount of early dark energy. Current bounds constrain it to be of the order $\alpha \sim 10$ or bigger \cite{doran_etal_2007} and here we use $\alpha = 10$. The homogeneous energy density and pressure of the scalar field $\phi$ are defined
in the usual way as \be \label{phi_bkg} \rho_{\phi} = \frac{\phi'^2}{2 a^2} + V(\phi)  \vv \,\,\, p_{\phi} = \frac{\phi'^2}{2 a^2} - V(\phi)  \vv \,\,\, w_{\phi} = \frac{p_{\phi}}{\rho_{\phi}} \pp \ee
We choose the value of $\gamma = -\beta/\alpha$ such that
we obtain the correct present dark energy fraction according to the relation \cite{wetterich_1995, amendola_2000},
\be
\Omega_h(t_0) = \frac{\gamma(t_0) m_\nu(t_0)}{16 eV} \,\,\,\,\, , \,\,\,\,\,\, \gamma(t_0) = -\frac{\beta(t_0)}{\alpha} \pp
\ee 

Finally, we can express the conservation equations for dark energy and growing
neutrinos as follows \cite{wetterich_1995, amendola_2000}: 
\bea \label{cons_phi} \rho_{\phi}' = -3 {\cal H} (1 + w_\phi) \rho_{\phi} +
\beta \phi' (1-3 w_{\nu}) \rho_{\nu} \vv \\
\label{cons_gr} \rho_{\nu}' = -3 {\cal H} (1 + w_{\nu}) \rho_{\nu} - \beta \phi' (1-3 w_{\nu}) \rho_{\nu}
\pp \eea The sum of the energy momentum tensors for neutrinos and dark energy is conserved,
 but not the separate parts. We neglect a possible cosmon coupling
to Cold Dark Matter (CDM), so that $\label{cons_cdm} \rho_c' = -3 {\cal H} \rho_c $.

\begin{figure}[ht]
\begin{center}
\includegraphics[width=85mm,angle=0.]{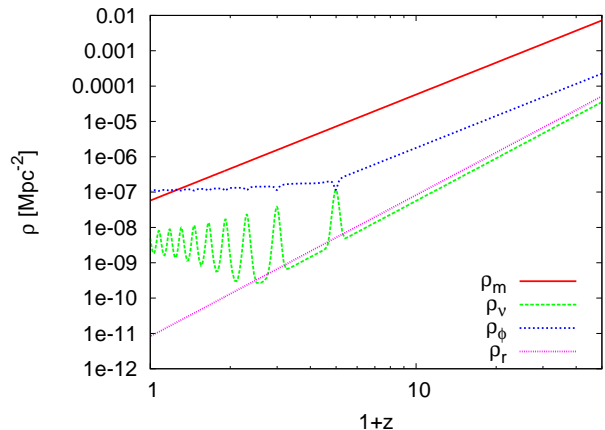}
\end{center}
\caption{Background evolution for fixed cosmon-neutrino coupling. Energy densities of cold dark matter (solid, red), neutrinos (long-dashed, green), dark energy (dot-dashed, blue) and photons (short dashed, black) are plotted vs redshift, for $\beta = -275$, $\alpha = 10$, $m_\nu(t_0) = 0.48$ eV.}
\label{fig_1_const}
\vspace{0.5cm}
\end{figure}

Given the potential (\ref{pot_def}), the evolution equations for the different species can be numerically integrated, providing the background
evolution shown in Fig.\ref{fig_1_const} for the model with constant $\beta$. The initial
pattern is a typical early dark energy model, since neutrinos are still relativistic and almost
massless, with $p_\nu = \rho_\nu/3$, so that the coupling term in eqs.(\ref{kg}), (\ref{cons_phi}), (\ref{cons_gr}) vanishes. Dark
energy is still subdominant and falls into the attractor provided by the
exponential potential (see \cite{wetterich_1995, amendola_2000, copeland_etal_1998} for
details). For $z \gsim 5$ it tracks the dominant background component with an early dark energy fraction $\Omega_h = n/\alpha^2$ and $n = 3(4)$ for the matter (radiation) dominated era.

\begin{figure}[ht]
\begin{center}
\includegraphics[width=85mm,angle=0.]{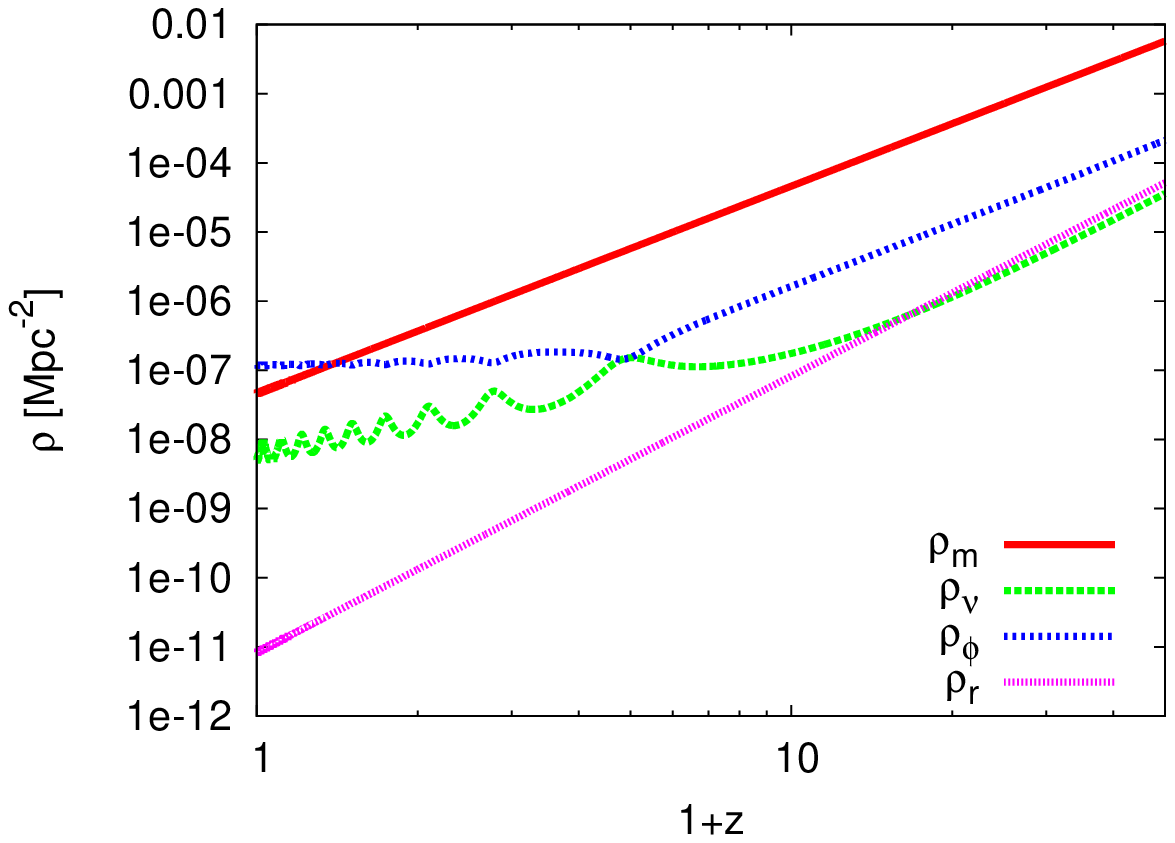}
\end{center}
\caption{Background evolution for varying cosmon-neutrino coupling. Energy densities of cold dark matter (solid, red), neutrinos (long-dashed, green), dark energy (dot-dashed, blue) and photons (short dashed, black) are plotted vs redshift, forgrowing coupling, $\alpha = 10$, $m_\nu(t_0) = 0.46$ eV.}
\label{fig_1_var}
\vspace{0.5cm}
\end{figure}

As the mass of the neutrinos increases with time, the coupling term $\sim \beta \rho_\nu$ in the evolution equation for the cosmon (\ref{kg}) starts to play a significant role as soon as neutrinos become nonrelativistic. This cosmological ``trigger event'' kicks $\phi$ out of the attractor. For $z < 5$ the value of the cosmon stays almost constant, and the frozen cosmon potential mimics a cosmological constant. In Fig.{\ref{fig_1_const}} this is visible in the modified behavior of $\rho_\nu$ and $\rho_\phi$ for $z < 5$. Subsequently, small decaying oscillations characterize the $\phi - \nu$ coupled fluid
and also $\rho_\nu$ reaches an almost constant value. For constant $\beta$ the increase of $m_\nu \sim a^3$ cancels the dilution of the neutrino number $n_\nu \sim a^{-3}$. This feature is altered for the growing coupling models (Fig.{\ref{fig_1_var}}). The va\-lues of the energy densities
today are in agreement with observations, once the precise crossing time for the end of the scaling 
solution has been fixed by an appropriate choice of parameters. At
present the neutrinos are still subdominant with respect to CDM, though in the future
they will take the lead.

\section{Linear and nonlinear evolution of neutrino fluctuations} \label{nl}

Once neutrinos become nonrelativistic ($z_{NR} \sim 5$), they start feeling the attractive force stronger than gravity mediated by the cosmon field. Neutrino perturbations rapidly grow and become nonlinear at a redshift $z_\text{nl} \sim 1-2$ \cite{mota_etal_2008}. Subsequently they might form stable lumps, whose solutions have been described in \cite{brouzakis_etal_2007,bernardini_etal_2009}. This scenario poses a challenge for our understanding of the spectrum of CMB anisotropies in the range of angular momenta $l \lsim 200$. In dependence on the size of the gravitational potential generated by the neutrino lumps one may find a strong influence on the ISW effect: photons falling into a potential well may come out with a different energy if the size of the gravitational potential has changed in the meantime. Details depend on the size and time evolution of the gravitational potential at various scales $k$.

In the relevant range of $k$ and $z$ the linear fluctuation equations for the neutrinos and the cosmon break down. Any realistic estimate for the size of the modifications of the CMB-spectra needs an estimate of the neutrino induced gravitational potential in the nonlinear regime. An extrapolation of the linear fluctuation equations \cite{franca_etal_2009} would give a large impact on the CMB-anisotropies which would rule out the models discussed in this paper. We will see, however, that this effect is overestimated by many orders of magnitude. In this paper we attempt a first estimate of the neutrino induced gravitational potential, based on extrapolations relying on the linear equations for small $k < 10^{-4}$ h/Mpc and nonlinear results for $k > 2\cdot10^{-2}$ h/Mpc. 

A summary of our findings is shown in Fig.\ref{fig:grav_pot_of_k}, where we display various approximations to the neutrino-induced gravitational potential $\Phi_\nu(k)$ for the present cosmological time ($z = 0$). Here $\Phi_\nu(k)$ is related in the Newtonian approximation to the neutrino density contrast $\Delta_\nu$
\be \label{eq:Phi_k_def} \Phi_\nu(k,z) = \frac{\rho_\nu(z) \Delta_\nu(k,z)}{2k^2(1+z)^2} \vv \ee
where the factor $(1+z)^{-2} = a^2(z)$ arises from the relation between comoving wave number $k$ and physical momentum $q = k/a$. The neutrino density contrast in Fourier space $\Delta_\nu(k)$ is obtained from the two point function or power spectrum $P_\nu(k) = V\exvalue{|\delta_\nu(k)|^2}$ as 
\be \label{eq:Delta_def} \Delta^2(k) = \frac{k^3 P_\nu(k)}{2 \pi^2} \vv \ee
where 
\be \delta_\nu(k) = V^{-1} \int_V d^3x \delta_\nu(x) e^{-i{\mathbf k}{\mathbf x}} \,\,\,,\, \delta_\nu(x) = \frac{\rho_\nu(x)}{\bar\rho_\nu} - 1 \pp \ee
In this figure, we have plotted the linear estimate (rising curve on the left) together with simple non-linear scenarios (falling curves on the right) and interpolations. Origin and details of these estimates will be given later in the text.

Unfortunately, the effects on the CMB spectrum are dominated by the gravitational potential for scales near the horizon, $k \approx 3\cdot10^{-4}$ h/Mpc. This is the region where the interpolation suffers from large uncertainties. As an overall picture we may state that interesting modifications of the CMB spectrum for $l \lsim 60$ occur if the neutrino induced gravitational potential reaches a value $\Phi_\nu(k)$ of a few times $10^{-6}$. For much larger values of $\Phi_\nu$ the models are ruled out, whereas for much smaller values the effects on the CMB become insignificant.

For small enough $k$ and large enough $z$ the linear fluctuation equations remain valid. For our system of coupled fluids they can be found in ref. \cite{mota_etal_2008}. The estimate of the scales and redshifts where the linear equations break down is not easy. There is an obvious breakdown when the neutrino overdensity $\Delta_\nu$ reaches unity. In practice, the critical value $\Delta_\nu^c$ for the breakdown of the linear evolution may be considerably smaller. A second important quantity is the product $\beta\,\delta\phi$, where $\delta\phi$ denotes the fluctuation of the cosmon field around its cosmological background value. For local values $|\beta\,\delta\phi| \approx 1$ the local neutrino masses differ strongly from the cosmological background value. The linear approximation assumes implicitly that the neutrino masses are given by the background value. It breaks down if this does no longer hold for a substantial fraction of the neutrinos. This extends to the effective value of the cosmon neutrino coupling $\beta$.

The hardest problem in this context concerns the issue of ``backreaction'' of ``small scale fluctuations'' well inside the horizon on ``large scale fluctuations'' close to the horizon. Evaluated for a given local neutrino lump the nonlinear effects arising for $|\beta\,\delta\phi| \gsim 1$ occur already for rather large $z$. Subsequently, the neutrino mass within a virialized lump may differ from the cosmological value outside the lump. As an effect, the fluctuations on larger length scales may ``see'' a modified effective neutrino mass and a modified $\phi$-dependence of $\bar{m}_\nu(\phi)$. This backreaction of fluctuations with larger $k$ on the evolution of fluctuations with smaller $k$ tells us that the bound of validity for the linear fluctuations at a given scale $k$ may not only depend on quantities evaluated at this scale.

A local value $\beta \delta \phi^{(l)} \approx 1$ for a lump of size $l$ corresponds to a much smaller value of the cosmological value $\beta \delta \phi(k)$ with $k = \pi a /l $. 
The averaging over the horizon volume contained in the Fourier transform leads to a suppression factor $\gamma_c$ which is given roughly by the total volume of all lumps
with size $\sim l$ divided by the horizon volume $\delta \phi (k) \approx \gamma_c \delta \phi_l (l = \pi a /k)$. In \cite{wintergerst_etal_2009} this suppression factor has
been estimated as $\gamma_c \approx 10^{-3}$. Thus linear perturbation theory could break down already for rather small values $\beta \delta \phi \approx 10^{-3}$. An explicit analysis of higher order terms in sect.\ref{sec:cosmon} shows that this is indeed the case.

On the opposite side for large $k$ and small $z$ the linear equations have clearly become meaningless. In order to understand the evolution of single neutrino lumps on scales well within the horizon one can solve numerically \cite{wintergerst_etal_2009} the Navier-Stokes equations in the Newtonian approximation. In an expanding universe with comoving coordinates $x$ they read in position space
\bea \label{eq:com_ns1} \delta_{\nu}' &=& -\scalprod{{\mathbf v}_{\nu}}{\mathbf \nabla}\delta_{\nu} - (1 + \delta_{\nu})\scalprod{\mathbf \nabla}{{\mathbf v}_{\nu}} \vv \\
\label{eq:com_ns2} {\mathbf v}_{\nu}' &=& -\left({\cal H} - \beta\phi'\right)\,\mathbf{v_{\nu}} - \left(\scalprod{\mathbf{v}_{\nu}}{\mathbf \nabla}\right)\mathbf{v}_{\nu} \nn \\
                                      &&\quad + {\bf \nabla} (\Phi_{\nu} + \beta\,\delta\phi) \vv \\
\label{eq:com_poisson} \Delta\delta\phi &=& -\beta\,a^2\,\delta_{\nu}\bar\rho_{\nu} \vv \\
\label{eq:com_grav_pot} \Delta\Phi_{\nu} &=& -\frac{a^2}{2}\,\delta_{\nu}\bar\rho_{\nu} \pp \eea
Here $\bar\rho_\nu$ is the background neutrino energy density and $\delta_\nu \equiv \delta \rho_\nu /\bar\rho_\nu$ is the relative neutrino density perturbation. The vector $\bf{v}_\nu$ is the velocity for neutrinos. More precisely, it describes the peculiar comoving velocities - it vanishes for neutrinos with constant comoving coordinates. The evolution of the velocities is driven by the gradients of the gravitational potential and cosmon field, with the usual Hubble damping and quadratic term arising from particle number conservation. The velocity dependent term $\beta\phi'\mathbf{v}_{\nu}$ in eq.(\ref{eq:com_ns2}) is not present in the standard Navier Stokes equations. It accounts for momentum conservation, reflecting the fact that the neutrino mass changes in time as $m'/m = -\beta \phi'$.
Equation (\ref{eq:com_grav_pot}) is the Poisson equation for the gravitational potential that we have indexed as $\Phi_{\nu}$ to clarify that it only comprises the neutrino contribution. Correspondingly, eq.(\ref{eq:com_poisson}) determines the local scalar potential $\delta\phi$. In this equation we have neglected the contribution from the cosmon potential $U(\phi)$. For the growing coupling model eq.(\ref{eq:com_poisson}) receives an additional contribution $\sim \partial \beta / \partial \phi$ that
we describe in sec.\ref{sec:cosmon}. We further note that the gravitational potential in position space as defined in eq.\,(\ref{eq:com_grav_pot}) is not the Fourier conjugate of $\Phi_\nu(k,z)$ in eq.\,(\ref{eq:Phi_k_def}), due to the definition of $\Delta(k)$ in eq.\,(\ref{eq:Delta_def}). For a detailed discussion of the underlying equations and the numerical solution techniques we refer to \cite{wintergerst_etal_2009}.

In ref. \cite{wintergerst_etal_2009} it was found that neutrino lumps virialize at a redshift $z_{vir} \approx 1.2$ for $m_\nu(t_0) = 2.1$ eV and a typical size $R_f = 14$ Mpc of the lump. Here we define the scale $R_f$ roughly as the scale where the gravitational potential reaches about one half of its value in the inner region of the lumps at the time of virialization. For single lumps it seems reasonable that the gravitational potential at the length scale $R_f$ does not change substantially for $z < z_{vir}$, since the details of what happens in the core of the lump should not matter for gravity outside.

The extrapolation from the gravitational potential of a single lump to the average cosmological gravitational potential at the corresponding scale $k$ is difficult. Since the latter involves a Fourier transform over distributions of lumps of different sizes, an estimate of the neutrino induced gravitational potential $\Phi_\nu(k)$ for strongly nonlinear scales has to assume some estimate of the number and distribution in size and mass of the virialized neutrino lumps. At the present time this estimate involves important uncertainties \cite{wintergerst_etal_2009}. Nevertheless, some lessons on the scale dependence, $z$-dependence and absolute upper bounds for $\Phi_\nu(k)$ can be drawn. 
We will discuss this issue in more detail in the next section.

The most uncertain part of our estimate of the fluctuation effects concerns the interpolation between the linear fluctuations and the 
``individual lump regime''.
This is the regime where the dynamical behavior of a gas of neutrino lumps matters
and where backreaction effects could be important. This interpolating regime makes the dominant contribution to the ISW effect.

After these general cautionary remarks we may now turn back to Fig.\ref{fig:grav_pot_of_k} where we show
the neutrino induced cosmological gravitational potential $\Phi_\nu(k)$ at $z=0$ for the different regimes in $k$,
evaluated for the model with varying $\beta$.
The extrapolation of the linear perturbation equations (green long-dashed line) leads to a strong increase of $\Phi_\nu(k)$ even for scales somewhat outside the horizon, 
where in absence of the cosmon mediated attraction between inhomogeneities one would have $\Phi_\nu \sim k^{-2}$. As a consequence, the
linearly extrapolated $\Phi_\nu(k)$ hits the extreme bound (for which all neutrinos within the horizon are concentrated in one spot) for length scales 
close to the horizon. As $k$ increases it soon reaches values that correspond to the potential near the Schwarzschild radius of a black hole.
It is obvious that this is not meaningful. We have indicated by a solid dot on the dashed green line the value of $k$ for which the linear approximation to $\Delta_\nu(k)$ reaches the value $1/4$, with larger values of $\Delta_\nu$ for larger $k$. In other words, $k_{nl} = 1.71\cdot10^{-3}$ h Mpc$^{-1}$ (corresponding to $R_{nl} = h \pi/k = 2.6$ Gpc) marks the smallest length scale for which the final linear density perturbation obeys $\left|\Delta_\nu(k)\right| = 1/4$. One sees that neutrino fluctuations on scales only slightly smaller than the horizon, $R_H \sim 3$ Gpc/h, are already highly nonlinear.

In the region $k \ge 0.01$ we indicate in Fig.\ref{fig:grav_pot_of_k} the result of rather radical and oversimplified assumptions. We suppose that one quarter of all cosmic neutrinos is clumped in lumps of a given size. For each lump we take the density profile that has been found from the solution of the hydrodynamical nonlinear equations for individual lumps in \cite{wintergerst_etal_2009}. We display two different sizes of lumps, $R_f = 15$ Mpc and $150$ Mpc. We expect the true neutrino induced gravitational potential to be below these curves. The decrease of the curves for large $k$ is due to the finite ``effective width'' of the neutrino distribution within the lumps. Concentrating all neutrinos of the lump in the center yields the dotted straight lines in Fig.\ref{fig:grav_pot_of_k}.
The extreme bound for $\Phi_\nu(k)$ assumes that all neutrinos within our horizon are concentrated in one point. It is indicated by the black dashed line. Indeed, if all neutrinos within the horizon volume $V_\text{hor}$ are concentrated in one point one finds
\be \label{eq:single_lump_Phi} P_\nu(k) = V_\text{hor}\,,\,\,\Delta_\nu^2(k) = \frac{k^3 V_\text{hor}}{2\pi^2}\,,\,\,\Phi_\nu(k) = \frac{\rho_\nu V_\text{hor}^{1/2}}{2\sqrt{2}\pi k^{1/2}} \pp \ee
If all neutrinos are distributed in $N$ equal pointlike lumps, the gravitational potential (\ref{eq:single_lump_Phi}) is multiplied by a factor $N^{-1/2}$, and a further factor $f_\nu$ arises if only a fraction $f_\nu$ of the neutrinos within the horizon are distributed in such pointlike lumps.

In the absence of nonlinear effects in the cosmon sector a typical interpolation would follow the line of linear growth until it hits a critical value, for which we take the extrapolation of eq.(\ref{eq:single_lump_Phi}) with $\Delta_\nu(k,z) = \Delta_\text{crit}$. In Fig.\ref{fig:grav_pot_of_k} we indicate the value $\Delta_\text{crit} = 1/4$ by a dot. Even though one expects some smoothening in the region where the two curves match,
the resulting values of $\Phi_\nu(k)$ typically would exceed a value $10^{-5}$ in the matching region. The ISW-contribution
would in this case exceed the observational bounds. 

For the interpolation shown in Fig.\ref{fig:grav_pot_of_k} the increase of $\Phi_\nu(k)$ outside and close to the horizon is damped by backreaction effects. This interpolation joins rather smoothly with the nonlinear regime for individual neutrino lumps for $k> 2 \cdot 10^{-2}$h Mpc$^{-1}$. The large values of $\Phi_\nu(k)$ close to the dot are never reached. The upper one of the blue dotted curves leads to an observable ISW-effect, while for the lower one no substantial impact on the CMB is found. This demonstrates the sensitivity of a possible CMB-signal to the details of the interpolation.

\section{Gravitational potential of neutrino lumps} \label{sec:results}
\begin{figure}[ht]
\begin{center}
\includegraphics[width=85mm,angle=-0.]{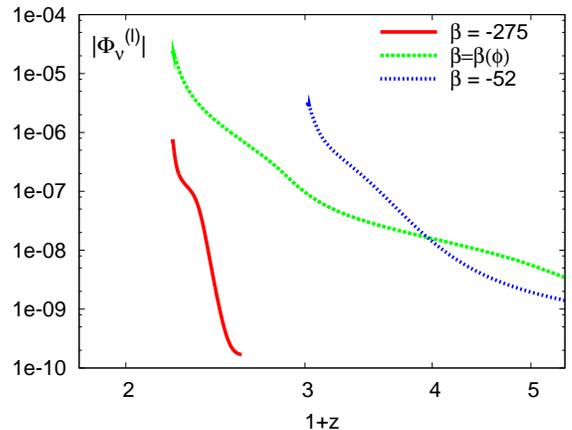}
\\
\end{center}
\caption{ Local neutrino gravitational potential $\Phi_{\nu}^{(l)}$ as a function of redshift for a fixed final effective gravitational radius $R_f = 14.3$ Mpc of the lump. We show models with fixed $\beta = -275$ (solid, red) $\beta = -52$ (dotted, blue) with corresponding neutrino masses $m_\nu(t_0) = 0.47$ or $2.11$ eV, respectively. The long-dashed, green curve refers to the growing coupling model with $m_\nu(t_0) = 0.46$ eV.}
\label{fig:grav_pot_vs_z}
\vspace{0.5cm}
\end{figure}

The local gravitational potential of a single lump can be directly accessed via a numerical integration of equations (\ref{eq:com_ns1})-(\ref{eq:com_grav_pot}) \cite{wintergerst_etal_2009}.
In Fig.\ref{fig:grav_pot_vs_z} we plot the time evolution of the gravitational potential of a single neutrino lump with final radius $R_f = 14.3$ Mpc for different growing neutrino models. 
Here $R_f$ is the distance from the center where the gravitational potential reaches one half of the value in the core at virialization. We show the local potential
(in position space) $\Phi_{\nu}^{(l)}(R_f)$.
In either model, the collapse of the lump and the growing neutrino mass lead to an increase of the gravitational potential. The $\phi$-dependence of $m_\nu$ is responsible for the observed small oscillations. Two features distinguish the particular models considered: 
\begin{itemize}
  \item[{\bf(i)}] {\bf Onset of growth:} In the model with a large constant $\beta = -275$ the growth of the neutrino mass is very rapid. Due to the smaller mass $m_\nu(t_0) = 0.47$ eV, the neutrinos become nonrelativistic later ($z_{NR} \sim 2$) than in the models with constant $\beta = -52$ and large neutrino mass $m_\nu(t_0) = 2.1$ eV, for which one finds $z_\text{NR} \sim 5$. For the growing coupling model one has $z_{NR} \sim 3$. For smaller values of $z_\text{NR}$ the growth of structure sets in later. For the red solid curve ($\beta = -275$) the neglection of pressure terms becomes justified only for $z < 2$. 
  \item[{\bf(ii)}] {\bf Time until virialization:} On the other hand, the stronger coupling leads to a substantial increase of the growth rate in the model with $\beta = -275$. Hence the time it takes from approximate linearity to virialization is much shorter in this model ($\Delta{z} \sim 0.4$) as compared to $\Delta{z} \sim 3$ for the model with constant $\beta = -52$ and large neutrino mass $m_\nu(t_0) = 2.1$ eV. This time is even longer for the varying coupling model.
\end{itemize}
The combination of both effects leads to a rather similar redshift of virialization $z_\text{vir}$ for the different models. For the two models under investigation we expect virialized structures around redshift $z_\text{vir} \sim 1.2$.

\begin{figure}[ht]
\begin{center}
\includegraphics[width=85mm,angle=-0.]{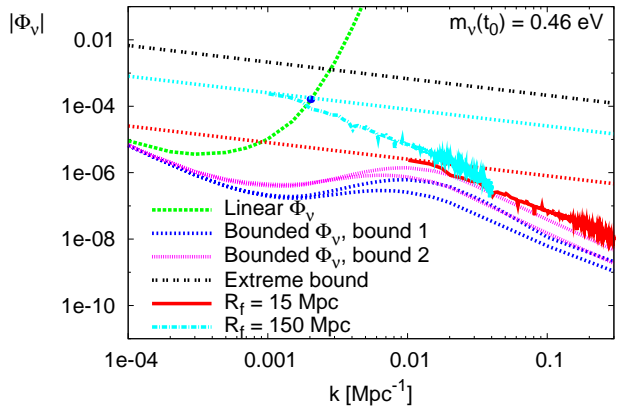}
\end{center}
\caption{Neutrino induced gravitational potential as a function of comoving wave number $k$ for $z = 0$. 
We show results for the growing coupling model with present average neutrino mass $m_\nu(t_0) = 0.46$ eV. 
The red and light blue lines are inferred from the nonlinear evolution. The solid lines indicate simple scenarios where one quarter of all neutrinos are clumped in extended lumps with the same size of $15$ (red) and $150$ Mpc (blue), respectively. The dashed lines are obtained from lumps concentrated in one point of corresponding mass. 
The behavior of $\Phi_\nu$ according to the linear approximation is shown by the long-dashed, green line. 
The dot indicates the value of $k$ for which the linear extrapolation of $\Delta_\nu(k)$ reaches the value $1/4$. 
The dot-dashed black line shows the extreme bound (\ref{eq:single_lump_Phi}). 
The interpolating dotted lines correspond to the bounded linear gravitational potential with $\eta_{crit} = 0.001$ and $\eta_{crit} = 0.01$, and $\Delta_{crit} = 0.1$ and $\Delta_{crit} = 1/4$, respectively, see details in sec.(\ref{sec:matching}). 
}
\label{fig:grav_pot_of_k}
\vspace{0.5cm}
\end{figure}

For an estimate of the cosmological impact of neutrino lumps one needs the average cosmological value of the gravitational potential at a given scale, expressed as the Fourier component $\Phi_{\nu}(k)$. By distributing lumps of given sizes in a cosmological volume, we have obtained rough estimates of this quantity as a function of the comoving scale $k$. We refer to \cite{wintergerst_etal_2009} for a detailed description of the procedure. 
As a rough approximation to these non-linear gravitational potentials in the range 
$0.03$ Mpc$^{-1} < k < 0.3 $ Mpc$^{-1}$ after virialization we find
\be \label{fit_alpha} \Phi_\nu(k,z)=c(z)k^{-\alpha}   \vv \,\,\,\, \alpha \approx -1.6 \pp 
\ee
For the rather crude nonlinear scenarios shown in Fig.\ref{fig:grav_pot_of_k} we obtain 
$ c(z_{vir}) \approx 1.2 \times 10^{-8} $, with $k$ in units of Mpc$^{-1}$.

The curves in Fig.\ref{fig:grav_pot_of_k} reflect the result of the rather radical assumption that all lumps form without substantial disturbance from other lumps in a situation close to spherical infall. Furthermore, the cosmological distribution of such lumps is strongly oversimplified by the assumption that a fraction $f_\nu = 1/4$ of the cosmic neutrinos is found in nonlinear lumps, and that all lumps are of equal size. We expect several effects to reduce the corresponding value of the neutrino induced gravitational potential $\Phi_\nu(k)$. A first substantial reduction arises if scattering and merging of smaller lumps plays a dominant role for the formation of larger lumps, rather than the roughly spherical infall for the formation of single lumps. This holds, in particular, if one attempts to extrapolate the estimate of the nonlinear $\Phi_\nu(k)$ to scales $k \lsim 6.3\cdot10^{-3}$ h/Mpc where the Newtonian approximation breaks down. Furthermore, backreaction effects typically slow down the growth of larger size neutrino structures,
again reducing the cosmological value $|\Phi_\nu(k)|$ as $k$ decreases (cf. the interpolated curve in Fig.\ref{fig:grav_pot_of_k}).

Finally, the neutrino-induced gravitational potential after virialization grows $\sim m_\nu(z)$. This follows if we assume that the neutrino number distribution $n_\nu(r)$, as a function of physical distance $r$ from the center of the lump, does not change any further after virialization. For the nonlinear estimates we have neglected possible backreaction effects and use the cosmological neutrino mass $\bar{m}_\nu(z)$ according to the background evolution (\ref{beta_phi}), (\ref{eq:nu_mass}). This typically leads to an increase of $\Phi_\nu(k)$ by a factor $(1+z_\text{vir})^3 \approx 10$ after virialization. The true value of the gravitational potential of neutrino lumps should rather use the average value of the neutrino mass in the lumps which may be substantially smaller than the cosmological value. In contrast, the lower two interpolated curves effectively stop the growth of $m_\nu$ within the lump once nonlinear effects become important. After virialization the time evolution of $\Phi_\nu(k)$ arises then only from the rescaling between comoving and physical momenta and a factor $a^{-3/2}$ from cosmological averaging (cf. sect.(\ref{sec:matching})).
This discussion reveals quite large potential errors in our nonlinear estimates which may rather be interpreted as approximate upper bounds. A reduction of $\Phi_\nu(k)$ by a factor $10$ or more seems quite reasonable, especially for small $k$. 

\begin{figure}[ht]
\begin{center}
\includegraphics[width=85mm,angle=0.]{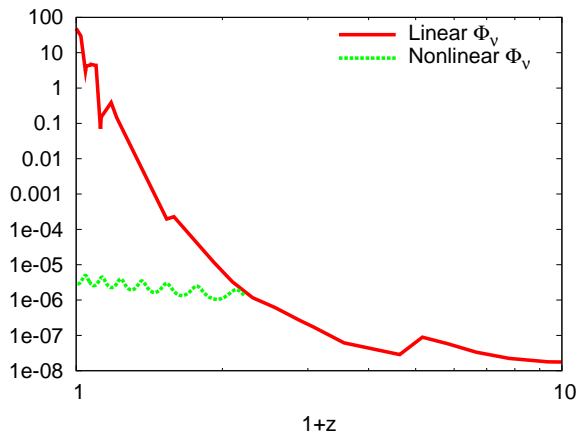}
\end{center}
\caption{Breakdown of the linear approximation for the gravitational potential for a physical wave number $q = k/a = 6.3\cdot10^{-3}$ h/Mpc. For the growing coupling model we show the redshift dependence of the linear approximation of $\Phi_\nu$ as the solid red line. It exceeds the rather extreme upper bound for $\Phi_\nu$ from 
a nonlinear estimate for final lumps of radius $R_f = 15$ Mpc which is given by the horizontal (wiggled) green line for redshifts after virialization.} 
\label{fig:phys_grav_pot}
\vspace{0.5cm}
\end{figure}

It is also worth to understand the evolution and breakdown of linear perturbation theory in more detail. In the linear approximation the gravitational potential is given by the expression
\be \label{eq:grav_pot_lin} \Phi_{\nu} \equiv \frac{a^2}{2k^2} \rho_\nu \left(\Delta_\nu + 3\frac{ {\cal H} }{k}v_\nu\right) \pp \ee
The first term denotes the density contribution and should match the nonlinear density contribution after virialization, whereas the second term, the ``velocity contribution'' approximately vanishes after virialization. In Fig.\ref{fig:phys_grav_pot} we plot the linear evolution for $\phi_\nu(q)$, evaluated at the physical scale $q=k/a = 6.3\cdot10^{-3}$ h/Mpc.
This should never exceed upper bounds obtained from the nonlinear evolution. In Fig.\ref{fig:phys_grav_pot} we compare this linear approximation with the crudely oversimplified assumption that only neutrinos in randomly distributed single lumps of final size $R_f = \pi/q$ contribute, and that a fraction $f_\nu^{(l)} = 0.25 $ of the total number of neutrinos is bound in such lumps. This results in the almost horizontal line in the left part of Fig.\ref{fig:phys_grav_pot}. We may
regard the corresponding green line as an upper bound for $\Phi_\nu$ at this scale. What is clear at this stage already is the huge overestimate of $\Phi_\nu$ which would result from the extrapolation of the linear approximation to the density contribution to $\Phi_\nu$ (shown in Fig.\ref{fig:phys_grav_pot} as the solid red line). 

\begin{figure}[ht]
\begin{center}
\includegraphics[width=85mm,angle=0.]{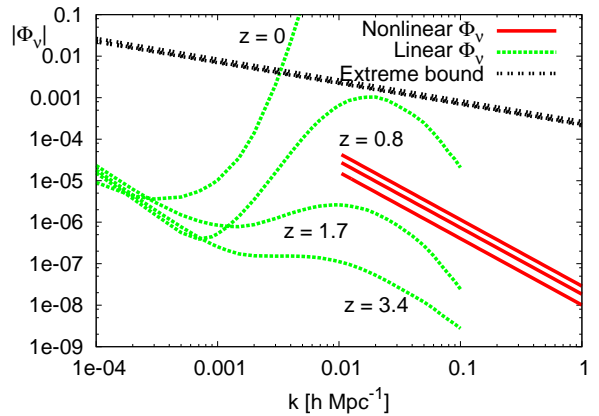}\\
\end{center}
\caption{
Neutrino induced gravitational potential vs. comoving wavenumber $k$ for the varying coupling model for redshifts $z = 0$, $z = 0.8$, $z = 1.7$, $z = 3.4$. Green lines show the result of linear perturbation theory, red lines the nonlinear estimates for $z = 0, 0.8, 1.7$ for a distribution of lumps with size $15$ Mpc.}
\label{fig:grav_pot_scale_comb}
\vspace{0.5cm}
\end{figure}

In Fig.\ref{fig:grav_pot_scale_comb}, we show for the growing coupling model the linear approximation to $\Phi_\nu(k)$
for various redshifts. This is confronted with randomly distributed nonlinear lumps of given size, represented as a power law approximation to the (scattered) nonlinear curves in Fig.\ref{fig:grav_pot_of_k}. We also indicate the extreme bound where all neutrinos are concentrated in one point. For $z = 3.4$ the linear approximation remains valid for the entire range of $k$. For
$z=z_b=1.7$ a range of $k$ between $0.02-0.06$ h Mpc$^{-1}$ reaches the nonlinear estimate. (Actually, the growth is somewhat
faster in the nonlinear treatment such that the nonlinear estimate is reached a bit earlier).
We will argue in the next section that the linear approximation breaks down for all wave numbers $k < 0.06$ h Mpc$^{-1}$ for $z<z_b$.
If one would continue the linear approximation for $z = 0.8$ or $z = 0$ the range of $k$ for which the nonlinear estimate is reached would gradually extend towards
the present horizon.
The interpolation shown in Fig.\,\ref{fig:grav_pot_of_k} corresponds to a freezing of the fluctuations for $k < 0.06$ h Mpc$^{-1}$ for $z<z_b$.

\section{Scalar potential of neutrino lumps and backreaction} \label{sec:cosmon}
The crucial ingredient in the analysis of neutrino fluctuations for a large neutrino-cosmon coupling $\beta$ is the local scalar potential \cite{wetterich_1995, peccei_etal_1987, ellis_etal_1989, wetterich_2003_jcap}. In parallel to Newton's gravitational potential it determines the strength of the attractive force between neutrinos, being stronger than gravity by a factor $2\beta^2$. Furthermore, the local scalar potential is directly related to the local deviation of the cosmon field from its cosmological average value. The local cosmon field determines the local neutrino mass. Possible backreaction effects of an effective neutrino mass in the lumps being smaller than the cosmological average are directly related to the strength of the local scalar potential. An investigation of the breakdown of linear perturbation theory may therefore concentrate on the evaluation of the size of non-linear effects for the values of the local cosmon potential and the corresponding cosmological average.

A neutrino lump not only generates a gravitational potential in its vicinity, but also a local scalar potential which is associated to a local deviation $\delta \phi$ of the cosmon field from its cosmological average value. The force on neutrinos exerted by the local scalar potential is enhanced by the coupling $\beta$, cf. eq. (\ref{eq:com_ns2}), such that the quantity to be compared with the gravitational potential is $\beta \delta \phi$. As long as eq. (\ref{eq:com_poisson}) remains a valid approximation, one has the simple relation $\delta \phi = 2 \beta \Phi_\nu$, cf. eq. (\ref{eq:com_grav_pot}). One arrives at a simple proportionality between the scalar potential and the neutrino induced gravitational potential 
\be
\beta \delta \phi = 2 \beta^2 \Phi_\nu \pp
\ee
This proportionality holds both for the local potential $\delta \phi_l$ and the cosmological averaged potential $\delta \phi(k)$.
Since the factor $2 \beta^2$ is large, for example $2 \beta^2 \sim 1.5 \times 10^{5}$ for the model with constant $\beta$ and $m_\nu(t_0) = 0.48$eV, the scalar potential $\beta \delta \phi$ reaches values close to one when the gravitational potential is still very small.
For values of $k$ within the horizon one can directly infer the scalar potential from Fig's \ref{fig:grav_pot_of_k} - \ref{fig:grav_pot_scale_comb} by multiplying $\Phi_\nu$ with a factor $2 \beta^2$.
We can relate the cosmological value of the scalar potential to the neutrino overdensity,
\be
\beta \delta \phi(k) = \frac{\beta^2 \Delta_\nu(k) \bar{\rho}_\nu a^2}{k^2} = \frac{3 \beta^2 \Omega_\nu(t_0) H_0^2}{k^2} \Delta_\nu(k) \vv
\ee
where the second identity holds for the present epoch $a = 1$, with $H_0$ the present Hubble parameter and $\Omega_\nu(t_0)$ given by eq.(\ref{omeganu}). We find that $\beta \delta \phi$ exceeds $\Delta_\nu$ for $k \lsim 0.01 $ Mpc$^{-1}$ (model with constant $\beta = -275$). In this range strong nonlinearities
in the cosmon fluctuations are expected even before $\Delta_\nu$ comes close to one.

As we have discussed in sec.\ref{nl}, local values of the scalar potential $\beta \delta \phi_l$ close to one can lead to important differences between a gas of neutrino lumps and a fluid of unclumped neutrinos. This is related to the nonlinear effects called ``backreaction''.
In presence of these nonlinear effects both the cosmological field equations for the background fields and the fluctuation equations for the modes with large wavelength are modified.
One possibly important effect concerns the relation between the cosmological average of the neutrino mass and the cosmological average of the cosmon field.
The average neutrino energy density, which influences $H$ and the evolution of neutrino overdensities, is given by 
\be
\bar{\rho}_\nu = \exvalue{m_\nu n_\nu} \equiv \langle m_\nu\rangle_n n_\nu \vv
\ee
where $n_\nu$ is the neutrino number density. The brackets indicate an averaging of the local value over the horizon and $\exvalue{}_n$
is the ``number-weighted average''. Since $m_\nu(\phi)$ is a nonlinear function of $\phi$, its number weighted average does not equal 
$m_\nu(\exvalue{\phi})$ as given by eq.(\ref{beta_phi}), with $\exvalue{\phi} = \phi$ the homogeneous cosmon ``background field''. Typically,
the local value of $\phi$ inside a lump is smaller than the cosmological average such that the neutrino mass inside the lump
is smaller than $m_\nu(\phi)$. If a large part of the neutrinos is concentrated in lumps this leads to a ``mass reduction factor'' $r_m$,
\be
r_m = \frac{\exvalue{m_\nu}_n}{m_\nu(\phi)} < 1 \pp
\ee

A second effect concerns the effective cosmon-neutrino coupling. In eq.(\ref{kg}) (and similarly in eq.(\ref{eq:com_poisson}) for the long wavelength fluctuations), one should modify the ``neutrino force'' $F_\nu = a^2 \beta \rho_\nu$ (we neglect pressure terms) by
\be
F_\nu = -a^2 \frac{\partial}{\partial \phi}\exvalue{m_\nu}_n n_\nu \pp
\ee
In consequence, one has to replace $\beta$ by an averaged interaction strength $\exvalue{\beta}$ as defined by 
\be
\exvalue{\beta} = r_\beta \beta \,\,\,\,\,   ,    \,\,\,\,\, r_\beta = \left( \left.\frac{\partial \ln\exvalue{m_\nu}_n}{\partial \phi} \right/ \frac{\partial \ln m_\nu(\phi)}{\partial \phi} \right) \pp
\ee
In view of the nonlinearities in the field equations it is well conceivable that the neutrino  mass inside a stabilized lump depends only
very mildly on the cosmological average field $\phi$. In this case one expects a strong reduction of the cosmon mediated attractive force between neutrino lumps, $r_\beta \ll 1$. This may have rather dramatic consequences for the growth of
fluctuations with long wavelength. A substantially reduced growth rate could effectively stop the increase of $\Delta_\nu(k)$ and $\Phi_\nu(k)$ for small $k$ as soon as the smaller scale lumps form and most neutrinos are in lumps for which the neutrino mass is insensitive to the average cosmon field $\phi$. 
Whenever $1-r_m$ or $1-r_\beta$ reaches values of the order one (say $1/4$) linear perturbation theory breaks down.
Finally,
also the potential derivative in eq.(\ref{kg}) has to be replaced by $\exvalue{\partial U / \partial \phi}$.

The breakdown of linear perturbation theory for the long-wavelength modes can be seen directly by an estimate of the relative size of the nonlinear terms in the evolution equations. As an example, we consider the model with varying $\beta$ and estimate the size of the nonlinear terms involving $\partial \beta / \partial \phi$. For a $\phi-$dependent $\beta$ eq.(\ref{eq:com_poisson}) is replaced by
\be
\Delta \delta \phi = \tilde{S}_L + \tilde{S}_{nl} \vv
\label{eq:totdphi}
\ee
\bea
\tilde{S}_L &=& -a^2 \bar{\rho_\nu} \left[ \beta(\phi) \delta_\nu + \frac{\partial \beta}{\partial \phi} \delta \phi \right] \nonumber
\\ &=& \tilde{S}_{L1} + \tilde{S}_{L2} \vv  \label{eq:sl12}\\
\tilde{S}_{nl} &=& - a^2 \bar{\rho}_\nu \left[ (\beta (\phi + \delta \phi) - \beta(\phi))(1 + \delta_\nu) 
- \frac{\partial \beta}{\partial \phi} \delta \phi \right] \nonumber \\
&\approx& -a^2 \bar{\rho_\nu} \frac{\partial \beta}{\partial \phi} \delta \phi \delta_\nu  \vv \label{eq:snl}
\eea
where we have neglected in the last line higher powers of $\delta \phi$. We observe that $\delta \phi$ has the opposite sign of $\delta_\nu$ while $\beta$ and $\partial \beta / \partial \phi = -\beta^2$ are both negative. Thus in the linear approximation the second term $\tilde{S}_{L2} \sim \delta \phi$ partially cancels the first term $\tilde{S}_{L1} \sim \delta_\nu$. Since the common prefactors do not matter for a comparison of the relative size of the different terms, we define $\tilde {S}=-a^2 \bar{\rho}_{\nu}\beta S$.

To evaluate the magnitude of the linear and nonlinear terms in Fourier space, we calculate the power 
spectrum of both terms to the first non trivial order. 
In Fourier space, the two terms $S_{L1}$ (we consider only this term for simplicity) and $S_{nl}$ can be written as
\begin{eqnarray}
S_{L1}&=&\quad N\int\delta_{\nu}(k)e^{i{\mathbf k}{\mathbf x}}d^{3}k \vv \\
S_{nl}&=&\quad \frac{\partial \ln \beta}{\partial\phi} N^{2}\int\delta_{\nu}(k_{1})\delta\phi( k_{2})e^{i({\mathbf k}_{1}+{\mathbf k}_{2}){\mathbf x}}d^{3}k_{1}d^{3}k_{2} \nonumber \vv
\end{eqnarray}
 where $N=V/(2\pi)^3$ is the Fourier prefactor.  At any given wavelength $k$, the two
terms contribute therefore as \begin{eqnarray}
N^{-1} S_{L1}=\quad &  & \delta_{\nu}(k)\nonumber\\
N^{-1} S_{nl}=\quad &  & \frac{\partial \ln \beta}{\partial\phi} N\int\delta_{\nu}(k_{1})\delta\phi(k-k_{1})d^{3}k_{1} \pp \nonumber
\end{eqnarray}

We want to see for which range of redshifts and $k$ the non-linear term gives a contribution similar in size to the linear terms. The linear approximation will clearly break down if this is the case. For this purpose we can take at a given $z$ the linear approximation to $\delta_\nu(k)$ and $\delta\phi(k)$ and evaluate the non-linear contribution in this approximation. Only if the non-linear contribution in this approximation remains small as compared to the linear one we can trust linear perturbation theory. Since by definition of the homogeneous background the cosmological average of the linear fluctuations $\delta_\nu(k)$ and $\delta\phi(k)$ vanishes 
we have to investigate the two-point function of the fluctuations of $\delta\phi$.

The variance of the linear term is proportional to 
\begin{equation}
G_{L1}(k,k')\equiv \langle\delta_{\nu}(k)\delta_{\nu}(k')\rangle=\frac{(2\pi)^{3}}{V^2}\delta^3(k+k')P_{\delta\delta}(k) \vv
\end{equation}
where  $V$ is the survey volume and $P_{\delta\delta}\equiv V|\delta_{\nu}(k)|^2$ is the power spectrum.
Integrating over $k'$ we have\begin{equation}
I_{L1}\equiv \frac{V^2}{(2\pi)^{3}}\int G_{L1}d^{3}k'=P_{\delta\delta}(k) \pp
\label{il1}
\end{equation}
In the Gaussian approximation one finds for the non-linear contribution to the two-point function to $\delta\phi$ a similar expression (we suppress
momentarily the $\nu$ subscript from $\delta_{\nu}$ and write momentum arguments as indices for clarity)
\begin{eqnarray}
G_{nl}(k,k')&=& (\frac{\partial \ln \beta}{\partial\phi})^2 N^2\int\langle\delta_{k_{1}}\delta\phi_{k_{2}}\delta_{k_{3}}\delta\phi_{k_{4}}\rangle\\
&\times&\delta^3(k-k_{1}-k_{2})\delta^{3}(k'-k_{3}-k_{4})\prod_{i=1}^{4}d^{3}k_{i} \pp \nonumber
\end{eqnarray}

Now the fourth-order term $\langle \delta_{k_{1}}\delta\phi_{k_{2}}\delta_{k_{3}}\delta\phi_{k_{4}} \rangle$ can be decomposed into
products of second order terms (since linear perturbations are supposed
to be Gaussian):\begin{eqnarray}
\langle\delta_{k_{1}}\delta\phi_{k_{2}}&\delta_{k_{3}}&\delta\phi_{k_{4}}\rangle = \langle\delta_{k_{1}}\delta\phi_{k_{2}}\rangle\langle\delta_{k_{3}}\delta\phi_{k_{4}}\rangle  \\
&+& \langle\delta_{k_{1}}\delta{}_{k_{3}}\rangle\langle\delta\phi_{k_{2}}\delta\phi_{k_{4}}\rangle +
\langle\delta_{k_{1}}\delta\phi_{k_{4}}\rangle\langle\delta\phi_{k_{2}}\delta{}_{k_{3}}\rangle
\pp \nonumber
\end{eqnarray}
Since $\langle\delta_{k_{1}}\delta\phi_{k_{2}}\rangle=
\frac{(2\pi)^{3}}{V^2}\delta^{3}(k_{1}+k_{2})P_{\nu\phi}$
and similarly for the other pairs, we obtain, successively integrating over $k_2,k_4,k'$,
\begin{eqnarray}
I_{nl} &\equiv& \frac{V^2}{(2\pi)^{3}}\int G_{nl}d^{3}k' =  \nonumber\\
&& (\frac{\partial \ln \beta}{\partial\phi})^2(2\pi)^{-3}
\int d^3 k_1 [P_{\nu\nu}(k_{1})P_{\phi\phi}(k-k_{1}) \nonumber\\
&&+ P_{\nu\phi}(k_{1})P_{\nu\phi}(k-k_{1})]
\end{eqnarray}
(a term proportional to $\delta^{3}(k)$ has been discarded since
it vanishes for all $k\not=0$). The position space variance is related to the power spectrum
$P(k)$ by
\begin{equation}
\sigma_{R}^{2}=\langle\delta(x)\delta(x+r)\rangle_{R}=\frac{1}{2\pi^2}\int P(k){W_{R}}^{2}(k)k^{2}dk\label{eq:s1}
\end{equation}
where $W_{R}(k)$ is the window function of a spherical cell of size
$R$ in Fourier space. Eq. (\ref{il1}) shows that  our definition of $I_{L1}$ is related to a directly observable 
quantity and similarly for the non-linear contribution $I_{nl}$. We see then that all factors containing the survey volume drops out in these quantities. (The output of CAMB gives the dimensionless quantities
$\Delta_{k}=\delta_{k}(k^{3} V/2\pi^{2})^{1/2}$ and the analogous expression for $ \delta\phi_k$.)
We next use the fact that $P_{\nu\nu}(k)$ and $P_{\nu\phi}(k)$ depend only on $|k|$ and that
 \begin{eqnarray}
I_{nl} & = & \frac{(\partial \ln \beta/\partial\phi)^2}{(2\pi)^{2}}\int_{0}^{\infty}k_{1}^{2}dk_{1}\int_{-1}^{1}d\mu \nonumber\\
 & \times & [P_{\nu\nu}(k_{1})P_{\phi\phi}(\sqrt{k^{2}+k_{1}^{2}-2kk_{1}\mu})\nonumber\\
&+& P_{\nu\phi}(k_{1})P_{\nu\phi}(\sqrt{k^{2}+k_{1}^{2}-2kk_{1}\mu})] \pp
\end{eqnarray}
 For small
$k$ the non-linear term becomes $k$-independent. We may now replace $P_{\nu\nu},P_{\nu\phi}$ by
$\Delta_{\nu\nu},\Delta_{\nu\phi}$,
\begin{equation}
P_{\alpha\beta} = 2\pi^2 \Delta^2_{\alpha\beta}k^{-3} \vv
\end{equation}
and extract $\Delta_{\alpha\beta}(k)$ from a CMB-code as CAMB. 

In the Newtonian approximation one has
\begin{eqnarray}
P_{\nu\phi}(k)&=&\frac{\beta^2a^2\bar{\rho}_{\nu}}{k^2}P_{\nu\nu}(k)=\frac{8\pi^2\beta}{a^2\bar{\rho}_\nu k}\Phi_{\nu}^2(k)\\
P_{\phi\phi}(k)&=&\frac{\beta^4a^4\bar{\rho}_{\nu}^2}{k^4}P_{\nu\nu}(k)=\frac{8\pi^2\beta^2}{ k^3}\Phi_{\nu}^2(k) \pp
\end{eqnarray}
We can therefore express $I_{nl}$ in terms of the cosmological neutrino induced Newtonian potential $\Phi_{\nu}$
and $P_{\nu\nu}$
 \begin{eqnarray}
I_{nl} & = & 2\beta^4 \left(\frac{\partial \ln \beta}{\partial\phi} \right)^2 \int_{0}^{\infty}\frac{dk_{1}}{k_1}\int_{-1}^{1}d\mu \nonumber\\
 & \times & \Phi_{\nu}^2(k_1) P_{\nu\nu}(\sqrt{k^{2}+k_{1}^{2}-2kk_{1}\mu} \,\,)\nonumber\\
&+& \left[\frac{k_1^2}{k^{2}+k_{1}^{2}-2kk_{1}\mu}+\frac{k_1^4}{(k^{2}+k_{1}^{2}-2kk_{1}\mu)^2}\right]
\end{eqnarray}
to be compared with $I_{L1}(k)=P_{\nu\nu}(k)$. For the growing coupling with $(\partial\ln\beta/\partial\phi)^2=\beta^2$ the nonlinear term typically reaches the same order
as the linear term if $\Phi_\nu$ reaches a value $\beta^{-2}$ in a relevant momentum range.

\begin{figure}[ht]
\begin{center}
\includegraphics[width=85mm,angle=0.]{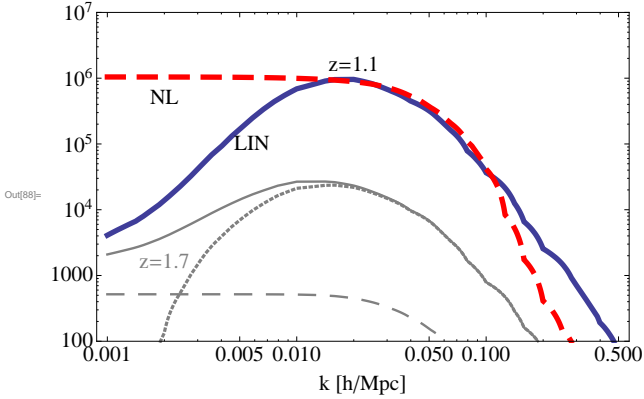}
\end{center}
\caption{Comparison of the linear (full lines) and nonlinear  (dashed lines) terms $I^{1/2}_{L1},I^{1/2}_{nl}$ at two epochs, $z_b=1.1$ (thick colored lines) and $z_b=1.7$ (thin gray lines). The dotted thin line is the combination of the two linear terms $L1$ and $L2$. For $z=1.1$ the linear approximation breaks down for all $k<0.2 h/$Mpc. } 
\label{fig:compare-lin-nonlin}
\vspace{0.5cm}
\end{figure}

We plot in Fig.\ref{fig:compare-lin-nonlin} the square root of the two terms $I_{L1},I_{nl}$
in arbitrary units at two epochs $z=1.1$ and $z=1.7$. (At these redshifts the quantity $\beta\delta\phi(k)$
reaches first a "threshold value" $\beta\delta\phi=0.01$ or $\beta\delta\phi=0.1$ for some momentum $k$. This will play a role for the modifications of the linear evolution discussed in the next section.) 
 As we can see, the two terms are comparable
already at $z\approx1.7$, and the non linear term equals or  dominates (for $k<0.1 h/$Mpc) the linear one
 at $z\approx1.1$. If we include the $S_{L2}$
term (see Eq. \ref{eq:sl12}), the linear term is further suppressed by the partial
cancellation and the non-linear contribution becomes relevant even earlier.

Further nonlinearities arise from deviations from the Gaussian approximation. The time evolution of the
fluctuations induces non-vanishing three-point functions and connected four-point functions (bi-spectrum and tri-spectrum) In turn, these also influence the evolution of the two-point function
\bea
\langle\delta\phi(k)&\delta&\phi(k')\rangle = \\ &&\frac{(2\pi)^3\delta^3(k+k')}{V^2}\frac{a^4\bar{\rho}^2\beta^4}{k^4}
(I_{L1}+I_{L2}+I_{nl}+...) \nonumber \vv
\eea
and add further nonlinearities to $I_{nl}$. We conclude that the cosmological fluctuations are already highly
non-linear at $z=1.1$. Nonlinearities presumably start to play an important role somewhere in the range $z=1.5-2$.

\section{Matching the linear and nonlinear evolution} \label{sec:matching}

Despite the present uncertainties in the estimate of the neutrino induced gravitational potential, one would like to get a first idea about
the possible impact of neutrino lumps on the CMB. In a sense, we would like to estimate the size of the neutrino lump induced nonlinear
Rees-Sciama effect. In order to implement the nonlinear features of the evolution of neutrino fluctuations
in a simple way in CMB-codes as CAMB or CMBEASY, we will work with simple bounds on the fluctuations
that should reflect the dominant features of the nonlinear effects. Of course, this can yield at best an idea about a possible size
of the effect. For every redshift $z$ these bounds have to match the linear evolution in its range of validity with the information and bounds available for
wave lengths in the nonlinear range.

In order to employ some reasonable procedure of matching the linear and the nonlinear gravitational potentials, it is useful to consider the time evolution in terms of the physical wavenumber $q$. The latter is related to the comoving scale $k$ via $k = aq$. In the linear regime for $z > z_\text{nl}$ we obtain for the density contribution of the gravitational potential:
\be \label{eq:grav_pot_phys} \Phi_\nu(q,z) = \frac{1}{2q^2}\,\rho_\nu(z)\Delta_\nu(q,z) \vv \ee
where $\Delta_\nu(q,z)$ changes according to the linear equations. 

We will assume that at a characteristic redshift $z_{nl}$ the non-linear effects become relevant for cosmological
averages and stop the further growth of the neutrino induced gravitational potential except for a change in the neutrino mass,
\be
\Phi_\nu(q,z)= \frac{m_\nu^{(l)}(z)}{m_{\nu}(z_{nl})} \Phi_\nu(q,z_{nl}) \pp
\label{grav_pot_39A}
\ee
This corresponds to a situation where the further infall of neutrinos into lumps of size $\pi/q$ has
a negligible effect on the gravitational potential. We take for $m_{\nu}(z_{nl})$ the cosmological neutrino mass, while $m_{\nu}^{(l)}(z)$ denotes the average neutrino mass in the lumps. 
For comoving momenta the gravitational potential (\ref{grav_pot_39A}) becomes 
\be
\Phi_\nu(k,z)= \left(\frac{a_{nl}}{a}\right)^{3/2} \frac{m_\nu^{(l)}(z)}{m_{\nu}(z_{nl})} \Phi_\nu \left(k \frac{a_{nl}}{a},z_{nl} \right) \vv
\ee
where the factor $a^{-3}$ in ${\Phi_\nu}^2(k)$ is due to the definitions of the green functions (it translates the $\delta-$function from q-space to k-space or, equivalently, accounts for the transition from physical to comoving volume). For $\Phi_\nu(q)\sim q^{- \alpha}$ this results in a time evolution 
\be
\Phi_\nu(k,z) \sim {m_\nu}^{(l)}(z) a^{\alpha-3/2}   \pp
\ee

We will discuss two different scenarios. For the first one (bound 1) the backreaction effects freeze the neutrino mass inside the lumps, $m_{\nu}^{(l)}(z) = m_\nu (z_{nl})$. For $\alpha = 2$ this results in 
\be
\Phi_\nu (k,z) = \sqrt{\frac{1+z_{nl}}{1+z}} \frac{\rho_\nu (z_{nl}) \Delta_\nu (z_{nl}(k))}{2 k^2 (1+z_{nl})^2} \vv
\label{grav_pot_39B}
\ee
while for $\alpha = 3/2$ (cf.\,eq.{\ref{fit_alpha}}) the first factor is cancelled and $\Phi_{\nu}(k,z)$ becomes independent of $z$.
The second scenario (bound 2) takes for ${m_\nu}^{(l)}(z)$ the cosmological neutrino mass, which multiplies the estimate of the first scenario for $\Phi_\nu$ by an oscillatory factor $m_{\nu}(z)/m_{\nu}(z_{nl})$. 
The numerical implementation of bound\,1 corresponds to $\Delta(k,z) = (a_b/a)^{7/2} \rho_\nu(z_b)/\rho_\nu(z) \Delta(k,z_b)$; bound\,2 is given by $\Delta(k,z) = (a_b/a)^{7/2} \rho_\nu(z_b)/\rho_\nu(z) \Delta(k,z_b) m_\nu(z)/m_\nu(z_b)$. The truth is probably somewhere in between.

At this stage we can devise simple strategies to implement the non-linear effects in a CMB-code. Of course, this can only be used for order of magnitude estimates of the effects of neutrino lumps, not for the computation of a detailed spectrum. Our general approach is to follow the linear evolution until its breakdown at redshift $z_{nl}$.
Subsequently we switch to the non-linear evolution according to eq.(\ref{grav_pot_39A}), employing either the scenario with constant $m_\nu^{(l)}(z)=m_\nu(z_{nl})$ or with varying neutrino mass $m_\nu(z)$ according to the cosmological solution. For the non-linear evolution we put the velocity contribution to $\Phi_\nu$ to zero. In general,
one may have different $z_{nl}$ for different $k-$modes. We therefore have to device a reasonable way to determine $z_{nl}$.

Consider first the case that the dominant non-linearity arises from the neutrino overdensity. In this case we choose a characteristic $\Delta_\nu^{crit}$ such that the linear theory breaks down for $\Delta_\nu(k) > \Delta_\nu^{crit}$.
We determine $z_{nl}(k)$ by the condition $\Delta_\nu(z_{nl}(k),k)=\Delta_\nu^{crit}$. Thus each mode is switched separately to the non-linear evolution when $\Delta_\nu(k)$ reaches the critical value. In practice we employ $\Delta_\nu^{crit} = 1/4$.

As we have discussed in sect.\ref{sec:cosmon}, neutrino overdensities close to one are not the only source of nonlinearity.
We should incorporate the large backreaction effect discussed in the preceding section. Its dominant role is to reduce 
the effective interaction $\exvalue{\beta}$ and the effective neutrino mass $\exvalue{m_\nu}_{n}$ ``seen'' by the long-distance modes.
Both effects reduce the growth of fluctuations as compared to the linear approximation.
One possibility would be the use of ``modified linear equations'' for which $m_{\nu}(\phi)$ and $\beta(\phi)$ are replaced by effective $k$-dependent 
functions $m_{\nu, \text{eff}}(\phi, k)$, $\beta_\text{eff}(\phi, k)$. An estimate of the effective functions requires an understanding
of the response of neutrino lumps of size $\pi a / k$ to variations of the ``outside cosmon field'' $\phi$, i.e. 
$\partial m_{\nu,\text{eff}}(\phi, k) / \partial \phi$.
We assume here that this response is much smaller than for the case of an unclumped neutrino fluid. In consequence, the reduction
of the growth rate for neutrino fluctuations with small $k$ is substantial, which justifies the use of eq.(\ref{grav_pot_39A}) for $z<z_{nl}$. We take here the crude approximation that the backreaction effects set in at some redshift $z_b$ for all
modes with $k<k_b$. Thus for all modes with $k<k_b$ we take $z_{nl}=z_b$ independently of $k$. (This holds unless $\Delta_\nu^{crit}$ has been reached even earlier for some $k$-modes. For these modes the ``$\Delta_\nu^{crit}$-trigger'' results in $z_{nl}(k) > z_b$.).

In practice we implement the backreaction effect in the CMB code by a ``backreaction bound'' for the growth of fluctuations, which acts in addition to the bound $\Delta_\nu(k)< \Delta_\text{crit}$. For this purpose we select a scale $k_b$ and follow the evolution of $\beta \delta\phi(k_b)$ until a critical value $\eta_\text{crit}$ is reached. At this value of $z = z_b$ we switch to the non-linear evolution for all modes with $k<k_b$. 
For the modes with $k>k_b$ the backreaction bound is not effective such that their growth is limited only by the condition $\Delta_\nu(k)<\Delta_\text{crit}$. For $k < k_b$, however, the end of the linear evolution and the approximate freezing of the growth of neutrino fluctuations according to eq.(\ref{grav_pot_39A}) is now set by the nonlinearity at the scale $k_b$, and no longer individually for every scale $k$. This reflects the overall picture that backreaction plays only a minor role for $k > k_b$, while it changes the behavior drastically for $k < k_b$. In other words, our procedure mirrors an effective $k$-dependent coupling $\beta(k,z)$ which is substantially smaller than $\beta$ for $z < z_b$ and $k < k_b$. We note that a strong backreaction effect would also affect the background evolution for $z < z_b$ by a modified effective $\beta$. We have not included such an effect yet. 
As long as the evolution of $\phi$ remains very slow also in presence of the backreaction, the overall cosmology for $z<z_b$,
for which dark energy behaves very close to a cosmological constant, will not be modified much.

\begin{figure}[h]
\begin{center}
\includegraphics[height=62mm,angle=0.]{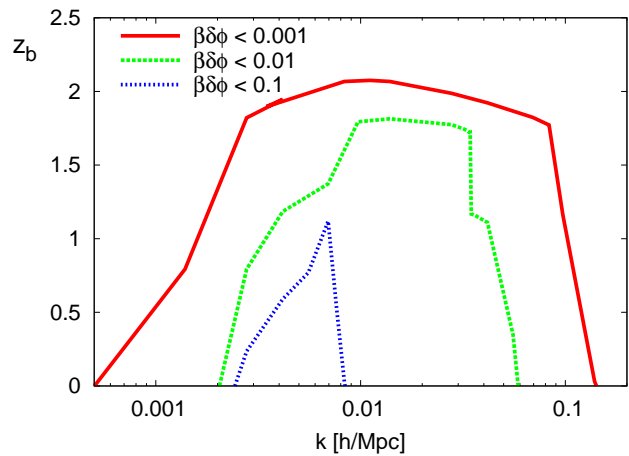}\\
\end{center}
\caption{Redshift $z_b$ at which the linear evolution of $\beta \delta \phi$ reaches the limit value $10^{-3}$, $10^{-2}$, $10^{-1}$ respectively, for the varying coupling model.}
\label{fig:znl_deltaphi}
\vspace{0.5cm}
\end{figure}

For an estimate of an appropriate choice of $k_b$ we plot in Fig.\ref{fig:znl_deltaphi} the redshift $z_b(k)$ for which the linear evolution of $\beta \delta \phi(k)$, constrained by the bound $\Delta_\nu(k)< \Delta_\text{crit} = 1 / 4$, reaches a given value $\beta \delta \phi(k) = 0.001, 0.01$ or $0.1$.
We observe that a whole range in $k$ reaches the given value of $\beta \delta \phi(k)$ almost simultaneously. This can
be understood by a glance at Fig.\ref{fig:grav_pot_scale_comb}, where for a whole range of $k$ the gravitational potential 
$\Phi_\nu(k)$ reaches a ``critical slope'' $\Phi_{\nu, \text{crit}} / k^2$ at $z=1.7$. (Since $\delta \phi(k) = 2 \beta^2 \Phi_\nu(k)$
we can use Fig.\ref{fig:grav_pot_scale_comb} for the time evolution of $\delta \phi(k)$, provided one rescales for every $z$
the units on the vertical axes by a factor $2 \beta^2(z)$.)
We choose $k_b$ at the right end of the plateau in Fig.\ref{fig:znl_deltaphi},
whereupon the precise choice within the plateau does not matter. In the varying coupling model we choose $z_b = 1.9 \,(1.7, 1.1)$ and $k_b = 5 \times10^{-2} \,(2.3 \times10^{-2}, 5 \times10^{-3})$ for $\eta_{crit} = 10^{-3} \,(10^{-2}, 10^{-1})$ respectively. In the constant coupling model $z_b = 1.4 \,(1)$ and $k_b = 3 \times10^{-2} \,(3.3 \times10^{-3})$ for $\eta_{crit} = 10^{-2} \,(10^{-1})$ respectively.
 For the choice of the value $\beta \delta \phi(k) = \eta_\text{crit}$, for which the backreaction stops the growth of fluctuations with $k < k_b$, we recall that the cosmological value of $\beta \delta \phi(k)$ is related to the local value of the scalar potential of individual lumps with radius $R_f = l$ by $\beta \delta \phi (k) = \gamma_c \beta \delta \phi^{(l)}$ ($l = \pi a / k$) by the dilution factor $\gamma_c$ in the range $10^{-3} - 10^{-2}$.
Since the neutrino mass within the lumps differs strongly from the cosmological neutrino mass for $|\beta \delta \phi^{(l)}(l)| = 1$, a typical value of $\eta_\text{crit}$ should be in the range of $\gamma_c$. We will display results for two choices, $\eta_\text{crit} = 0.01$ and $\eta_\text{crit} = 0.001$. Furthermore, the interpolation becomes very smooth if at the redshift where $\beta \delta \phi(k_b)$ reaches $\eta_\text{crit}$ also $\Delta_\nu(k_b)$ reaches $\Delta_\text{crit}$.
This is realized for the pairs ($\Delta_\text{crit} = 1/4, \eta_\text{crit} = 0.01$) and ($\Delta_\text{crit} = 10^{-1}, \eta_\text{crit} = 10^{-3}$),
that we will use for our estimate of the CMB anisotropies for the growing coupling model. 

\begin{figure}[ht]
\begin{center}
\includegraphics[height=60mm,angle=0.]{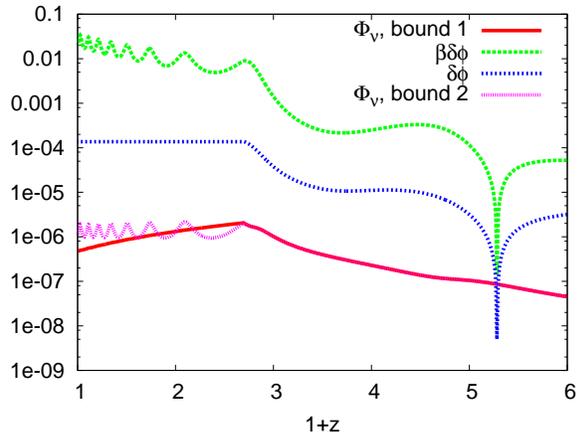}
\end{center}
\caption{Evolution of perturbations from the bounded linear code, for a scale $k = 0.005$ 1/Mpc. The bounds are set by 
$\Delta_\text{crit} = 1/4, \eta_\text{crit} = 0.01, k_b = 0.023, 1+ z_b = 2.7$. For this $k$ scale $\eta_\text{crit}$ is reached before $\Delta_\text{crit}$. Bound 1 and bound 2 correspond to the conditions given above.}
\label{fig:bounded_densities}
\vspace{0.5cm}
\end{figure}

A typical evolution of fluctuations according to our switch to the non-linear evolution is shown in Fig.\ref{fig:bounded_densities}.

\section{Impact of neutrino lumps on CMB}

\begin{figure}[ht]
\begin{center}
\includegraphics[height=62mm,angle=0.]{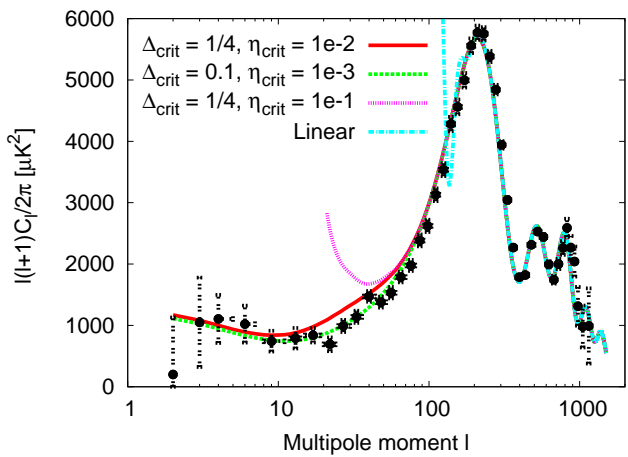}
\end{center}
\caption{Angular power spectrum of CMB anisotropies for the varying coupling model. The lines correspond to different approximations for the nonlinear neutrino fluctuations as explained in the text.}
\label{fig:cls_var}
\vspace{0.5cm}
\end{figure}

The impact of the neutrino spectrum on the CMB-anisotropies strongly depends on the evolution of the fluctuations in the nonlinear range.
We demonstrate this for the growing coupling model with a present average neutrino mass $m_\nu(t_0) = 0.46$eV in Fig.\ref{fig:cls_var}.
This figure shows the spectrum of anisotropies for various approximations:
\begin{enumerate}
\item Linear perturbation theory (blue dot-dashed line) shows drastic deviations from the observed spectrum for $l \le 200$. This result agrees with
\cite{franca_etal_2009}. As we have argued, linear perturbation theory is not appropriate, however.
\item Imposing the bound $\Delta_\nu(k) < 0.1$, together with the backreaction bound $\beta \delta \phi(k_b) < 10^{-3}$ for $k_b = 5 \times 10^{-2}$ and using the bound 1 for the evolution at $z<z_{nl}$, we find the green dashed line. The neutrino fluctuations leave almost no imprint on the spectrum. Deviations from the $\Lambda$CDM model result
primarily from the presence of early dark energy.
\item For a less restrictive bound $\Delta_\nu(k) < 0.25$, $\beta \delta \phi (k_b) < 10^{-2}$ the neutrino induced gravitational potential
is somewhat larger (cf. the upper vs the lower dotted blue line (bound 1) in Fig.\ref{fig:grav_pot_of_k})). Now a signal 
of the neutrino fluctuations is visible as an enhancement of the anisotropies in the range $10 \le l \le 80$,
as shown in the red solid line in Fig.\ref{fig:cls_var}.
This gives an indication of the size of the ISW - or nonlinear Rees-Sciama effect for a neutrino induced gravitational potential
of the size shown in Fig.\ref{fig:grav_pot_of_k}. The smoothness of this effect is a result of the particular type of bounds employed.
Other implementations of the nonlinear neutrino fluctuations can lead to an oscillatory behavior.
\item A much stronger ISW effect would result for an increased neutrino induced gravitational potential. The 
purple dotted line shows the spectrum for the bound $\beta \delta \phi (k_b) < 0.1$. Strong discrepancies
with observations would occur for $l \le 80$. In view of the estimate of the backreaction effect in sect.\ref{sec:matching}
this curve cannot be trusted, however. The employed linear approximation is invalid within the range of redshifts during which the large fluctuations responsible for the ISW-effect grow.
\end{enumerate}

For all curves we use the cosmological parameters shown in the first column of table (\ref{tab:param_comp}).
\begin{table}[ht]
\begin{tabular}{|c|c|c|}
\hline
\bf{growing coupling}&\bf{fixed coupling}\\
\toprule
$m_\nu(t_0) = 0.46$ eV & $m_\nu(t_0) = 0.48$ eV\\
\hline
$\alpha = 10$ & $\alpha = 10$\\
$\phi_t = -0.005$ & $\beta = -275$\\
\hline
$\Omega_{\phi0} = 0.67$ &  $\Omega_{\phi0} = 0.65$\\
$\Omega_{c0} = 0.26$ &  $\Omega_{c0} = 0.276$ \\
$\Omega_{\nu0} = 2.82\cdot10^{-2}$  & $\Omega_{\nu0} = 3.12\cdot10^{-2}$\\
$\Omega_{b0} = 4.14\cdot10^{-2}$ &  $\Omega_{b0} = 4.34\cdot10^{-2}$\\
$H_0 = 72.9$ & $H_0 = 71.3$ \\
\hline
\end{tabular}
\caption{Cosmological parameters for a varying $\beta(\phi)$ model and a constant $\beta$ model.}
\label{tab:param_comp}
\end{table}
These parameters are modified as compared to the ones for the $\Lambda$CDM model due to the presence of early dark energy.
For larger values of $\alpha$, resulting in a reduced early dark energy, they would
be closer to the $\Lambda$CDM model. We have not attempted an optimization of parameters in this paper.
Deviations of the shown spectrum from the $\Lambda$CDM model for $l>200$ can be reduced by an improved
parameter choice. Restrictions on models arising from the CMB spectrum for $l>200$ or the corresponding power spectrum
of matter fluctuations for $k > 3 \times 10^{-2}$ Mpc$^{-1}$ mainly concern a bound on the parameter $\alpha$ and are not addressed in this paper.

\begin{figure}[ht]
\begin{center}
\includegraphics[height=62mm,angle=0.]{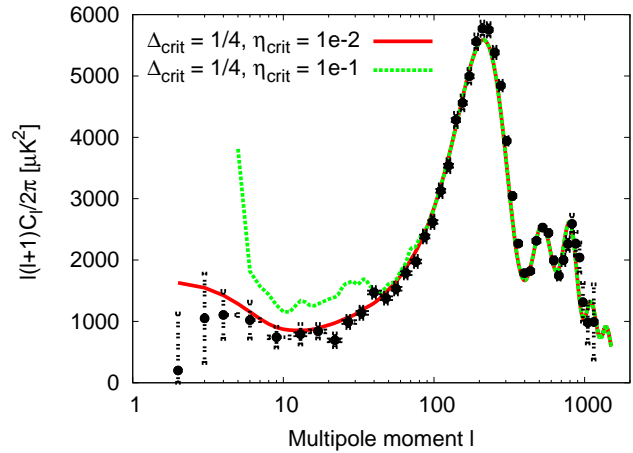}
\end{center}
\caption{Angular power spectrum of CMB anisotropies for the constant coupling model, $\beta = -275$.}
\label{fig:cls}
\vspace{0.5cm}
\end{figure}

The CMB-anisotropies are sensitive to the neutrino mass and the form of the neutrino-cosmon
coupling. In Fig.\ref{fig:cls} we plot results for similar bounds as in Fig.\ref{fig:cls_var},
but now for a fixed coupling $\beta = -275$. The cosmological parameters used for this figure are displayed in the second column of
Tab.\ref{tab:param_comp}).

\section{Conclusions} \label{conclusions}

We have attempted to estimate the size of the ISW-effect for the CMB spectrum which results from the formation of large scale neutrino lumps
in models of growing neutrino quintessence. Many of our general considerations also apply for other models for which the coupling between neutrinos and a light scalar field substantially exceeds gravitational strength, as for the mass varying neutrino models of \cite{fardon_etal_2004,bjaelde_etal_2007,brookfield_etal_2005}.
Our most central conclusion seems quite robust: linear perturbation theory cannot be trusted for an estimate of the CMB spectrum for these models. This often concerns already the onset of observable differences from the $\Lambda$CDM model, since the neutrino fluctuations which dominate the ISW-effect in the linear approximation actually correspond to a range of wave numbers $k$ for which the linear approximation is no longer valid.

For an estimate of the CMB-spectrum for $l < 100$ in growing neutrino quintessence one needs to solve a rather complex nonlinear problem, 
similar to the estimate of the nonlinear Rees-Sciama effect for $l > 1500$ arising from dark matter lumps. In contrast to the dark
matter lumps, N-body simulations or reliable analytic results for the neutrino-lump-induced gravitational potential $\Phi_\nu(k)$
in the relevant range $k \le 0.04$ h Mpc$^{-1}$ are not yet available. The challenge for such methods consists in an appropriate
treatment of a locally varying scalar field in the presence of possibly important backreaction effects.
Waiting for more accurate results for $\Phi_\nu(k)$ and its evolution with redshift, we attempt an estimate of the plausible overall size of the ISW effect by
imposing reasonable bounds on the growth of fluctuations.

For the varying coupling model we learn from Fig.\ref{fig:cls_var} that an observable ISW-effect for $l \le 100$ seems plausible,
perhaps dominating around $l \approx 40$.
It seems well conceivable that this effect includes oscillatory features, reflecting oscillations in the cosmological neutrino
energy density as visible in Fig.\ref{fig_1_var}. One may even speculate that this could be responsible for small
``enhancements'' and ``depressions'' in the observed spectrum for this range of $l$. The size and details
of the neutrino-induced ISW-effect depend on the particular model and its parameters. Unfortunately,
a detailed parameter estimate seems premature in view of the large uncertainties affecting the average
gravitational potential of the neutrino lumps.
Nevertheless, it may be possible to exclude certain ranges of models and parameters if it is plausible that the deviations from the linear behavior
in the relevant ranges of $k$ and $z$ are under control.

A general lesson concerns the presence of strong backreaction effects within growing neutrino quintessence. A gas
of neutrino lumps behaves differently from an approximately homogeneous fluid of neutrinos. The microscopic values
$m_\nu(\phi)$ and $\beta(\phi)$ of the neutrino mass and cosmon-neutrino coupling get replaced by an effective mass and coupling once the
neutrino lumps have become sufficiently nonlinear. This seems to be one of the first examples of a realistic large backreaction effect which modifies the
parameters of the homogeneous and isotropic background field equations by order one effects.
A self consistent treatment beyond linear perturbation theory will have to take them into account.
As a whole, the findings of this paper are encouraging for more detailed investigations of growing neutrino 
quintessence. It seems that such models could be consistent with observations for appropriate
ranges of parameters. Once the true evolution of the neutrino-induced cosmological gravitational potential $\Phi_\nu(k,z)$
is better constrained, interesting model tests become feasible through neutrino-induced modifications of the CMB-spectrum for $l \le 100$,
an enhanced correlation between matter and photon fluctuations or an enhanced large scale bulk flow.
The possible existence of large scale neutrino lumps could offer new surprises for cosmology. 

\begin{acknowledgments}
We thank David F. Mota for providing us with a first version of CAMB code including mass varying neutrinos. The work of NW is supported by the Humboldt Foundation.
\end{acknowledgments}

\section*{Appendix}

\section*{Linear perturbations within general growing neutrino models}

We recall here the linear perturbation equations for growing neutrino quintessence, following the work
presented in \cite{mota_etal_2008}, to which we refer for further details.
The evolution equations for linear perturbations (in Fourier space), in Newtonian gauge (in which the non diagonal metric perturbations are fixed to zero) \cite{kodama_sasaki_1984}, read for the growing neutrino scenario:
\bea \delta_{\phi} ' &=&  3 {\cal H} (w_{\phi} - c_{{{\phi}}}^2) \delta_{\phi}  \nonumber \\ &-& \beta (\phi) \phi' \frac{\rho_\nu}{\rho_{\phi}} \left[(1-3 w_\nu) \delta_{\phi} - (1-3 c_{{\nu}}^2) \delta_\nu \right]  \nonumber \\ &-&(1+ w_\phi)(k v_{\phi} + 3 {\Phi}') \nonumber \\& +& \frac{\rho_\nu}{\rho_{\phi}} (1-3 w_\nu) \left(\beta(\phi) \delta \phi' + \frac{d \beta(\phi)}{d \phi} \phi' \, \delta \phi  \right)  \vv \eea
\bea
 \delta_{\nu}' &=& 3 ({\cal H} - \beta(\phi) \phi') (w_{\nu} - c_{{\nu}}^2) \delta_{\nu}  \nonumber \\ &-& (1+w_{\nu})(k v_\nu + 3 {{\Phi}'}) - \beta(\phi) (1-3 w_{\nu}) \delta \phi'  \nonumber \\ &-& \frac{d \beta(\phi)}{d \phi} \phi' \delta \phi \, (1-3 w_\nu) \pp
\eea
The equations for the density contrasts $\delta_i(k) = \frac{1}{V}\int{\delta_i({\bf x}) exp(-i {\bf{k \cdot x}}) d^3x }$ (defined as the Fourier transformation of the local density perturbation $\delta_i({\bf x}) = \delta \rho_i(x)/\rho_i(x)$ over a volume $V$) involve the velocity perturbations, 
which evolve according to
\bea 
v_{\phi}' &=& -{\cal H}(1-3 w_{\phi}) v_{\phi} - \beta(\phi)\phi'(1-3 w_\nu) \frac{\rho_\nu}{\rho_{\phi}} v_{\phi}  \nonumber \\ &-& \frac{w_{\phi}'}{1+w_{\phi}} v_{\phi} + k c_{{{\phi}}}^2 \frac{\delta_{\phi}}{1+w_{\phi}} +  k { \Psi}  \nonumber \\ &-&  \frac{2}{3} \frac{w_{\phi}}{1+w_{\phi}} k \pi_{T_{\phi}} + k \beta(\phi) \delta \phi \frac{\rho_\nu}{\rho_{\phi}} \frac{1-3 w_\nu}{1+w_{\phi}}  \vv
\eea
\bea v_\nu ' &=& (1 - 3 w_\nu) (\beta(\phi) \phi'-{\cal H}) v_\nu - \frac{w'_{\nu}}{1+w_{\nu}} v_\nu  \nonumber \\ &+& k c_{{\nu}}^2 \frac{\delta_{\nu}}{1+w_{\nu}} + k {{\Psi}} - \frac{2}{3} k \frac{w_{\nu}}{1+w_{\nu}} \pi_{T \, \nu} \nonumber  \\ &-& k \beta(\phi) \delta \phi \frac{1-3 w_\nu}{1+w_\nu} \pp \eea

As usual, the gravitational potentials obey
\be \label{Phi} { \Phi} = \frac{a^2}{2 k^2 M^2} \left[\sum_\alpha \left(\delta \rho_\alpha + 3 \frac{{\cal H}}{k} \rho_\alpha(1+w_\alpha)v_\alpha \right) \right] \vv
\ee 
\be { \Psi} = -{ \Phi} - \frac{a^2}{k^2 M^2} \sum_\alpha w_\alpha \rho_\alpha \pi_{T\alpha} \vv \ee 
where $\pi_{T\alpha}$ is the anisotropic stress for the species $\alpha$ and the sound velocities are defined by $c_{i}^2 \equiv \delta p_i / \delta \rho_i$. The perturbed pressure for $\phi$ is 
\be \delta p_{\phi} = \frac{\phi'}{a^2} \delta \phi' - \frac{{ \Psi}}{a^2}\phi'^2 - U_\phi \delta \phi \ee 
and the anisotropic stress $\pi_{T_{\phi}} = 0$, as in uncoupled quintessence, since the coupling is treated as an external source in the Einstein equations. The linear perturbation of the cosmon, $\delta \phi $, is related to $v_{\phi}$ via \bea \delta \phi &=& \phi' v_{\phi} / k \vv \nonumber \\ \delta \phi ' &=& \frac{\phi' v_{\phi}'}{k} + \frac{1}{k} \left[ -2 {\cal H} \phi' - a^2 \frac{dU}{d \phi} \right. \\ && \left.  + a^2 \beta(\phi) (\rho_\nu - 3 p_\nu) \right] v_{\phi}  \nonumber \pp \eea
Note that $\delta \phi$ can equivalently be obtained as the solution of the perturbed Klein Gordon equation:
\bea \delta \phi '' &+& 2 {\cal H} \delta \phi' + \left(k^2 + a^2 \frac{d^2 U}{d \phi^2}\right) \delta \phi  - \phi'({ \Psi} ' - 3 { \Phi} ') \nonumber \\  &+& 2 a^2 \frac{d U}{d \phi} { \Psi} = - a^2 \left[-\beta(\phi) \rho_\nu \delta_\nu (1-3 c_{\nu}^2)  \right. \\ &-& \left. \frac{d \beta(\phi)}{d \phi} \delta \phi \rho_\nu (1 - 3 w_\nu) - 2 \beta(\phi)(\rho_\nu - 3 p_\nu) { \Psi} \right] \pp \nonumber \eea

The evolution of neutrinos requires solving the Boltzmann equation in the case in which an interaction between neutrinos and the cosmon is present \cite{ichiki_keum_2007}. The first order Boltzmann equation written in Newtonian gauge reads \cite{ma_bertschinger_1995}
\bea
\frac{\partial{\Psi_{ps}}}{\partial \tau} &+& i \frac{q}{\epsilon}({\bf k} \cdot {\bf n}) \Psi_{ps} + \frac{d\ln{f_0}}{d\ln{q}} 
\left[ -{ \Phi}' - i \frac{\epsilon}{q} ({\bf k} \cdot {\bf n}) { \Psi} \right] = \nonumber \\ &=& i \frac{q}{\epsilon}({\bf k} \cdot {\bf n}) k \frac{a^2 m^2_\nu}{q^2} \frac{\partial \ln{m_\nu}}{\partial \phi} \frac{d \ln{f_0}}{d \ln{q}} \delta \phi
\vv \eea
\be f(x^i, \tau, q, n_j) = f_0(q) \left[ 1 +\Psi_{ps}(x^i,\tau, q, n_j) \right] \vv \ee
${{\Psi}}$ and ${ \Phi}$ are the metric perturbations, $x^i$ the spatial coordinates, $\tau$ is the conformal time, ${\bf q} = a {\bf p} = q {\bf \hat{n}}$ is the comoving 3-momentum, $\epsilon = \epsilon(\phi) = \sqrt{q^2 +m_\nu(\phi)^2 a^2} $, $f$ is the phase space distribution and $f_0$ its zeroth-order term (Fermi-Dirac distribution).
\\
The Boltzmann hierarchy for neutrinos, obtained expanding the perturbation ${\Psi_{ps}}$ in a Legendre series can be written in Newtonian gauge as
\bea \Psi_{ps, 0}' &=& -\frac{q k}{\epsilon} \Psi_{ps, 1} + { \Phi}' \frac{d\ln{f_0}}{d \ln{q}} \vv  \\
 \Psi_{ps, 1}' &=& \frac{q k}{3 \epsilon} (\Psi_{ps, 0} - 2 \Psi_{ps, 2}) - \frac{\epsilon k}{3 q} { \Psi} \frac{d \ln{f_0}}{d \ln{q}} + \kappa \nonumber \vv \\
 \Psi'_{ps, l} &=& \frac{qk}{(2l+1) \epsilon} \left[l \Psi_{ps, l-1} - (l+1) \Psi_{ps, l+1} \right] \nonumber \,\,\,\, l\ge 2 \vv \eea
where \cite{ma_bertschinger_1995, ichiki_keum_2007}
\be \kappa =  - \frac{1}{3} \frac{q}{\epsilon}k \frac{a^2 m^2_\nu}{q^2} \frac{\partial \ln{m_\nu}}{\partial \phi} \frac{d \ln{f_0}}{d \ln{q}} \delta \phi \pp \ee
\\
This allows us to calculate the perturbed energy and pressure as well as the shear for neutrinos:
\bea
 &&\delta \rho_\nu = a^{-4} \int{q^2 f_0(q) \left[\epsilon(\phi) \Psi_{ps, 0} + \frac{\partial \epsilon}{\partial \phi} \delta \phi \right] dq d\Omega}  \vv \nonumber \\
 &&\delta p_\nu = \frac{a^{-4}}{3} \int{\frac{q^4}{\epsilon^2} f_0(q) \left[ \epsilon  \Psi_{ps, 0} - \frac{\partial \epsilon}{\partial \phi} \delta \phi \right]dq d\Omega}  \vv \nonumber \\
&& (\rho_\nu + p_\nu)\sigma_{\nu} =  \frac{8 \pi}{3} a^{-4} \int{q^2 dq \frac{q^2}{\epsilon} f_0(q) \Psi_{ps, 2}} 
 \pp \eea
The anisotropic stress is related to the shear via $\pi_{T_\nu} = \frac{3}{2 p_\nu}(\rho_\nu + p_\nu) \sigma $ and in our case 
\be \frac{\partial \epsilon}{\partial \phi} = \frac{a^2 m_\nu^2}{\epsilon} \frac{\partial \ln{m_\nu}}{\partial \phi} = - \beta(\phi) \frac{a^2 m_\nu^2(\phi)}{\epsilon(\phi)} \pp \ee
Note also that the unperturbed neutrino density and pressure read
\bea
\rho_\nu &=& a^{-4} \int{q^2 dq d\Omega {\epsilon(\phi)} f_0(q)}  \vv \\
p_\nu &=& \frac{1}{3} a^{-4} \int{q^2  dq d\Omega \frac{q^2}{\epsilon}(\phi) f_0(q)} \pp
\eea
\end{document}